\documentclass[11pt,a4,english]{article}
\usepackage[T1]{fontenc}
\usepackage[latin9]{inputenc}
\usepackage{geometry}
\usepackage{verbatim}
\usepackage{amsmath}
\usepackage{graphicx}
\usepackage{longtable}
\usepackage{setspace}
\usepackage[round,sort,authoryear]{natbib}
\usepackage{amsfonts}
\usepackage{graphicx}
\usepackage{babel}
\usepackage{amsthm}
\usepackage[toc,page]{appendix}
\usepackage{booktabs,caption}
\usepackage{enumitem}
\usepackage{amsfonts}
\usepackage{accents}
\usepackage{comment}
\usepackage{natbib}
\usepackage{bibentry}
\usepackage{amsmath}
\usepackage[colorlinks=true,linkcolor=blue, citecolor=blue, urlcolor = blue]{hyperref}
\usepackage{lmodern}
\usepackage{avant}
\usepackage[T1]{fontenc}
\usepackage[center]{titlesec}
\titleformat*{\section}{\bfseries\centering}
\titleformat*{\subsection}{\bfseries}
\titleformat*{\subsubsection}{\bfseries} 

\providecommand{\U}[1]{\protect\rule{.1in}{.1in}}
\newtheorem{theorem}{Theorem}

\newtheorem{definition}{Definition}
\newtheorem{example}{Example}

\newtheorem{remark}{Remark}

\newtheorem{assumption}{Assumption}

\usepackage{mathtools}

\makeatletter
\newcommand{\nextverbatimspread}[1]{%
  \def\verbatim@font{%
    \linespread{#1}\normalfont\ttfamily% Updated definition
    \gdef\verbatim@font{\normalfont\ttfamily}}% Revert to old definition
}
\makeatother

\DeclarePairedDelimiter{\floor}{\lfloor}{\rfloor}

\numberwithin{equation}{section}

\AtBeginDocument{}

\newif\ifshow 
\showfalse 

\ifshow
  \includecomment{wrap}
\else
  \excludecomment{wrap} % anything wrapped in a {wrap} environment will be excluded
\fi

\begin{document}

\title{ {\Large \textbf{Partial Sum Processes of Residual-Based and Wald-type Break-Point Statistics in Time Series Regression Models}\thanks{Article History: May 2020. December 2021. 

I would like to thank Jose Olmo, Tassos Magdalinos and Jean-Yves Pitarakis for guidance, support and continuous encouragement throughout the PhD programme. Financial support from the VC PhD studentship of the University of Southampton is gratefully acknowledged. The author declares no conflicts of interest. All remaining errors are mine.}
} }

\author{
\\
\textbf{Christis Katsouris}\footnote{Ph.D. Candidate, Department of Economics, University of Southampton, Southampton, SO17 1BJ, UK. \textit{E-mail Address}: \textcolor{blue}{C.Katsouris@soton.ac.uk}.}
\\
\small{Department of Economics, University of Southampton}
}

\date{}

%\date{January 31, 2022}

\maketitle
\vspace*{-1.4 em}
\begin{center}
\textbf{Abstract}
\end{center}
\vspace*{-3.8 em}
\begin{abstract}
We revisit classical asymptotics when testing for a structural break in linear regression models by obtaining  the limit theory of residual-based and Wald-type processes. First, we establish the Brownian bridge limiting distribution of these test statistics. Second, we study the asymptotic behaviour of the partial-sum processes in nonstationary (linear) time series regression models. Although, the particular comparisons of these two different modelling environments is done from the perspective of the partial-sum processes, it emphasizes that the presence of nuisance parameters can change the asymptotic behaviour of the functionals under consideration. Simulation experiments verify size distortions when testing for a break in nonstationary time series regressions which indicates that the normalized Brownian bridge limit cannot provide a suitable asymptotic approximation in this case. Further research is required to establish the cause of size distortions under the null hypothesis of parameter stability. 
\\

\\

\end{abstract}

%%-------------------------------------------------------------------------%%
%\newpage 

\section{INTRODUCTION}

Severe time series fluctuations manifesting as market exuberance are considered by econometricians as early warning signs of upcoming economic recessions which are not usually explained by common boom and bust cycles (see, \cite{greenwood2020predictable}, \cite{baron2021banking}  and \cite{katsouris2021sequential}). Furthermore, the economic aspects of prolonged economic policy uncertainty as well as as more recently pandemics can impact the robustness of parameter estimates due to increased model uncertainty. More precisely, such economic phenomena can appear in specification functions as structural breaks to parameter coefficients. Therefore, the identification and estimation of the true break-points can improve forecasts and reduce model uncertainty. Specifically, in the literature there is a plethora of statistical methodologies for structural break detection under different modelling environments  (see, \cite{bai1997estimating}, \cite{bai1998estimating} and \cite{pitarakis2004least} among others). In this paper we focus on the properties of the partial sum processes when constructing test statistics for testing the null hypothesis of no parameter instability in time series regression models under the assumption of stationary regressors vis-a-vis nonstationary regressors.

%%-------------------------------------------------------------------------%%
\newpage 

Although we do not introduce any new testing methodologies we review some important asymptotic theory results related to testing for structural breaks in time series regression models. More precisely, we study the asymptotic behaviour of partial-sum processes when these are employed to construct residual-based and Wald-type statistics in time series regression models. Our motivation for revisiting some of these aspects is to add on the discussion regarding the adequacy of finite-distribution approximations to large-sample theory of statistics. In particular, a common misconception when deriving asymptotic theory of test statistics and estimators is the fact that the large sample theory certainly provides good approximations to finite-sample results (see, \cite{chernoff1958asymptotic}). Lastly, we consider two different modelling frameworks that is a stationary time series regression model and a nonstationary time series regression. 

In terms of the structural break testing framework, we consider various examples which cover both the case of a known break-point as well as an unknown break-point, under parametric assumptions regarding the distribution of the innovations. Specifically, throughout this paper we assume that the innovations driving the error terms of both the stationary and nonstationary time series models are $\textit{i.i.d}$, with a known distribution, that is, are Gaussian distributed which implies that we are within the realm of parametric methods. Moreover, relaxing the particular assumption by assuming that distribution function of these innovations is considered as a nuisance parameter, requires to consider a semiparametric framework for estimation and inference purposes which is beyond the scope of this paper. Therefore, we have that the sequence of innovations, $\{ \epsilon_t  \}_{t=1}^{\infty}$ to be $\epsilon_t \sim_{ \textit{i.i.d} } \mathcal{N}(0, \sigma^2)$ which implies independence and homoscedasticity\footnote{Notice that these two assumptions can indeed be quite strong. For example weakly dependent and heteroscedastic innovations can reflect more accurately the properties of aggregate time series and therefore such assumptions can be also included via appropriate modifications.}. Furthermore, the assumption of stationary innovation sequences is also imposed to facilitate the limit theory and this property holds regardless of the time series properties of the regressors and regressand. 

In particular, we are interested in obtaining limit results of the following form
\begin{align}
\underset{ 0 \leq s \leq 1  }{ \mathsf{sup} } | \mathcal{C}_T(s) | \overset{ d }{ \to } \underset{ 0 \leq s \leq 1  }{ \mathsf{sup} } | W(s) | 
\end{align}
where $\mathcal{C}_T(s)$ is the test statistic and $W(s)$ is the standard Wiener process for some $s \in [0,1]$. However, the emphasis in this paper is the investigation of the main asymptotic theory aspects based of the regression model under consideration in terms of the properties of regressors. Therefore, the specific comparison allow us to focus on the implementation of the partial-sum processes when deriving asymptotic theory results for test statistics for these two classes of time series regression models, that is, the classical linear regression versus nonstationary time series regression. The investigation of the properties of partial-sum processes for the residuals of non-linear regression models such as ARMA, ARMAX, ARCH or GARCH models can be indeed quite fruitful, but we leave the particular considerations for future research. Related studies include the papers of \cite{kulperger2005high}, \cite{aue2006strong} and \cite{gronneberg2019partial}.  

All limits are taken as $T \to \infty$ where $T$ is the sample size. The symbol $"\Rightarrow"$ is used to denote the weak convergence of the associated probability measures as $T \to \infty$ (as defined in \cite{billingsley2013convergence}) which implies convergence of sequences of random vectors of continuous cadlag functions on $[0,1]$ within a Skorokhod topology. The symbol $\overset{d}{\to}$ denotes convergence in distribution and $\overset{p}{\to}$ denotes convergence in probability. The remainder of the paper is organized as follows. Section \ref{Section2} discusses some examples related to the use of partial-sum processes for residual-based statistics in linear regression models. Section \ref{Section3} discusses Wald-type statistics in linear regression models. Section \ref{Section4} discusses some aspects related to the asymptotic theory for the structural break testing framework in nonstationary time series regressions. Section \ref{Section5} concludes and discusses further aspects for research.

%%-------------------------------------------------------------------------%%
\newpage 

\section{RESIDUAL-BASED STATISTICS}
\label{Section2}

The use of partial-sum processes for model residuals when testing the null hypothesis of no structural break using residual-based statistics appears in various applications in the literature \citep{katsouris2021sequential}. Firstly, \cite{brown1975techniques} proposed the OLS-CUSUM test constructed based on cumulated sums of recursive residuals for testing for the presence of a single structural break in coefficients of the linear regression model (see, also \cite{kramer1988testing} and \cite{ploberger1992cusum}). Other residual-based statistics found in the literature include the OLS-MOSUM test proposed by \cite{chu1995mosum} which is constructed as sums of a fixed number of residuals that move across the whole sample. Therefore, in this case the statistic can be more sensitive in detecting parameter changes in comparison to the cumulated sums of recursive residuals. Furthermore, the literature evolved towards the construction of structural break statistics within an online monitoring framework which includes a window of fixed size (historical period) and an out-of-sample estimation window (monitoring period). In particular, \cite{chu1995moving} and \cite{kuan1994implementing}  proposed the Moving Estimates and Recursive Estimates statistics for parameter stability respectively. A unified framework has been then proposed by the seminal study of  \cite{chu1996monitoring} and also further examined by \cite{leisch2000monitoring} for the case of the generalized fluctuation test.

Some additional considerations include the study of the characteristics of structural change which includes the frequency of breaks in time series such as single vis-a-vis multiple break points (e.g.,   \cite{bai1997estimating} and \cite{bai1998estimating}) as well as the nature of structural change, which implies detecting structural breaks in the conditional mean vis-a-vis the conditional variance or higher moments of partial sum processes (e.g., \cite{horvath2001empirical}, \cite{kulperger2005high}, \cite{pitarakis2004least}, \cite{andreou2006monitoring, andreou2009structural}). Furthermore, an alternative asymptotic analysis of residual-based statistics is proposed by \cite{andreou2012alternative}. The current literature extensively studies structural changes in the mean and variance of regression coefficients, however less attention is given to the study of structural break testing due to smooth changes in the persistence properties of regressors as it is defined within the framework of local-to-unity for autoregressive models. Specifically, smooth transitions of stochastic processes from $I(0)$ to $I(1)$ and other non-stationarities can be employed for the detection of bubbles in financial markets (see, \cite{horvath2020sequential}). In this paper, our aim is to provide a discussion of the use of partial sum processes of residual-based and Wald-type statistics for these two different modelling environments.

\subsection{OLS-CUSUM test}

The OLS-CUSUM test statistic (see, \cite{kramer1988testing}) belongs to the class of residual based statistics (see, e.g., \cite{stock1994unit}) and it involves the partial sum processes of regression residuals based on the model under consideration. Following the literature we define a general class of regression residuals based on the partial sum process as proposed by \cite{kulperger2005high} given by Definition \ref{definition1} and Theorem \ref{Theorem1} below.    

\begin{definition}
\label{definition1}
The partial sum process of residuals is given by 
\begin{align*}
\widehat{ \mathcal{U} }_T (s) =  \sum_{t=1}^{ \floor{ Ts} } \widehat{u}_t , \ \ \text{for some} \ s \in [0,1].
\end{align*}
with the partial sum process of the corresponding innovations is similarly defined as
\begin{align*}
\mathcal{U}_T (s) =  \sum_{t=1}^{ \floor{ Ts} } u_t, \ \ \text{for some} \ s \in [0,1].
\end{align*}
\end{definition}

\begin{theorem}\label{Theorem1}
Suppose that $\sqrt{T}  \big| \hat{ \theta}_T - \theta \big| = \mathcal{O}_p(1)$ where $\hat{\theta}_T$ is the set of estimated model parameters and $\theta$ is in the interior of $\Theta$. If $\mathbb{E} \left( | u_0| \right) < \infty$, then 
\begin{align}
\underset{ s \in [0,1] }{  \mathsf{sup} } \ \frac{1}{ \sqrt{T} } \ \bigg| \bigg( \widehat{ \mathcal{U} }_T(s) - s \widehat{ \mathcal{U} }_T(s) \bigg) - \bigg(  \mathcal{U}_T(s) - s \mathcal{U}_T(s) \bigg) \bigg|  = o_p(1).
\end{align}
Then, the invariance principle for partial sums for an \textit{i.i.d} sequence $\big\{ u_t \big\}$ such that 
\begin{align}
\left\{ M_T (s) :=  \frac{ \mathcal{U}_T(s) - s \mathcal{U}_T(s)  }{ \sigma_u \sqrt{T} }, \ s \in [0,1] \right\}  \ \ \text{and} \ \  \left\{ \widehat{M}_T (s) :=  \frac{ \widehat{ \mathcal{U} }_T(s) - s \widehat{ \mathcal{U} }_T(s) }{ \sigma_u \sqrt{T} }, \ s \in [0,1] \right\} 
\end{align} 
where $\sigma_u^2 = \mathbb{E}\left( u^2 - \mu_u \right)^2$ and $\mu_u = \mathbb{E}\left( u_0 \right)$, implies that the functionals $M_T (s)$ and $\widehat{M}_T (s)$ both converge weakly in the Skorokhod topology $\mathcal{D}[0,1]$, to a Brownian bridge $\bigg\{ \mathcal{BB}(s) = W(s) - s W(1), \ s \in [0,1] \bigg\}$. 
\end{theorem}

\medskip

\begin{remark}
The functional $M_T (s)$ represents a partial-sum process and $W(s)$ is the corresponding standard Wiener process on the interval $[0,1]$ which preserves the continuous mapping theorem and convergence in a suitable functional space for any continuous function $g(.)$, such that $g\left( M_T (s) \right) \overset{\mathcal{D}}{\to} g \left( W(s) \right)$. Furthermore, both Definition \ref{definition1} and Theorem \ref{Theorem1} corresponds to partial-sum process of the residual sequence of a time series regression model with a general specification form, under the assumption of stationarity and ergodicity. Similarly, we can generalize these limit results to second order partial-sum processes for square residual sequences. More precisely, the first and second order partial-sum processes represent functionals of the OLS-CUSUM and OLS-CUSUM squared (see, \cite{deng2008limit}).  
\end{remark}

The proof of Theorem \ref{Theorem1} is omitted which demonstrates a weak invariance principle; a stronger version of Donsker's classical functional central limit theorem (see, \cite{kulperger2005high} and \citep{csorgHo2003donsker}). In particular, the weakly convergence of the asymptotic distribution of the OLS-CUSUM statistic for the classical regression model is studied by \cite{aue2013structural}. Related limit results can be found in the book of \cite{csorgo1997limit}. The development of the asymptotic theory for the residual-based and Wald-type statistics when testing for  structural breaks is based on the validity of Theorem \ref{Theorem1}. For the remaining of this section we consider some standard examples from the literature to demonstrate the use of the residual-based statistics and their corresponding limit results when conducting statistical inference.   

\medskip

\begin{example}
\label{Example1}
Consider the location model formulated as below 
\begin{align}
\label{location}
y_t = \mu_1 \mathbf{1} \{ t \leq k \}   + \mu_2 \mathbf{1} \{ t > k  \} + \epsilon_t, \ t =1,...,T  
\end{align}
where $\mu_1$ and $\mu_2$ are both deterministic and the break point $k = \floor{Ts}$ for some $s \in [0,1]$ is an unknown fixed fraction of the full sample. Suppose a parametric assumption on the error term holds, such that $\epsilon_t \sim_{ \textit{i.i.d} } \mathcal{N} (0,1)$ for $t = 1,...,T$ and the following moment conditions apply
\begin{align}
\textit{(i).} \ \  \mathbb{E} \left( \epsilon_t | \mathcal{F}_{t-1} \right) = 0 \ \ \ \ \ \textit{(ii).} \ \  \mathbb{E} \left( \epsilon^2_t | \mathcal{F}_{t-1} \right) = \sigma^2_{\epsilon} \ \ \ \ \ \textit{(iii).} \ \  \mathbb{E} \left( | \epsilon_t | \right)^{2 + d} < \infty,  \ \ d > 0.
\end{align}
Then, the OLS estimator $\widehat{\mu} = T^{-1} \sum_{t=1}^T y_t$ of $\mu$ is a $\sqrt{T}-$consistent estimator and asymptotically normal. Furthermore, based on the location model formulation given by \eqref{location} the objective is to conduct a two-sided test of the testing hypothesis $\mathbb{H}_0: \mu_1 = \mu_2$. Under the null hypothesis we obtain a consistent estimator $\hat{\sigma}^2_{\epsilon}$ of $\sigma^2_{\epsilon} < \infty$ such that $\hat{\sigma}^2_{\epsilon} \overset{  p }{ \to } \sigma^2_{\epsilon}$ as $T \to \infty$ as well as the OLS-residuals defined as  $\widehat{\epsilon}_t = \big( y_t - \widehat{\mu} \big)$.

%%-------------------------------------------------------------------------%%
\newpage 

Thus, we have that 
\begin{align}
\widehat{\epsilon}_t 
\equiv 
\big( y_t - \bar{y}_T  \big) 
=  
\mu + \epsilon_t - \frac{1}{T} \sum_{t=1}^T \left( \mu + \epsilon_t \right)
\end{align}
\end{example}
\vspace{-0.8em}
Based on Example \ref{Example1}, we consider that the following functional central limit theorem (FCLT) holds 
\begin{align}
\frac{1}{\sqrt{T} } \sum_{t=1}^{[Ts]} \epsilon_t \Rightarrow \sigma_{\epsilon} W(s), \ \ s \in [0,1].
\end{align}
which applies to the unobservable innovation terms of the above time series model that includes only a model intercept, where $W(.)$ is the standard Brownian motion such that $W(s) \sim N(0,s)$. For notational convenience we also denote with  $\sigma_{\epsilon} W(s) \equiv B(s)$ for some $s \in [0,1]$. 

Consider the standardized innovations defined as 
$\epsilon_t^o = \displaystyle \left(  \epsilon_t - \frac{1}{T} \sum_{t=1}^T \epsilon_t \right) \equiv \big( \epsilon_t -  \bar{\epsilon} \big)$ and suppose that $\epsilon_t$ is an $\textit{i.i.d}$ sequence of innovations. Then the centered partial sum process is defined as below
\begin{align}
M_T (s) = \frac{1}{\sqrt{T} } \sum_{t=1}^{[Ts]} \big( \epsilon_t -  \bar{\epsilon} \big), \ \ s \in [0,1].
\end{align}
and the corresponding residual centered partial sum process is defined as below
\begin{align}
\widehat{M}_T (s) = \frac{1}{\sqrt{T} } \sum_{t=1}^{[Ts]} \big( \widehat{\epsilon}_t - \bar{ \widehat{\epsilon} } \big),\ \ s \in [0,1].
\end{align}
Moreover, let $\widehat{\sigma}^2_m = \widehat{M}^2_T (1) / T$ to be the sample variance estimator of the above partial sum process. Then, 
\begin{align}
\underset{  0 \leq s \leq 1 }{ \mathsf{sup} } \frac{1}{ \sqrt{T} } \left|  \frac{ \widehat{M}_T (s) }{ \widehat{\sigma}^2_m  } - \frac{ M_T (s) }{ \sigma^2_m  } \right| = o_p(1)
\end{align}
which shows that the self-normalized centered partial sum process  $\big\{ \widehat{M}_T (s) / \widehat{\sigma}^2_m, \ 0 \leq s \leq 1 \big\}$ behaves as the residuals $\left\{ \widehat{\epsilon}_t \right\}_{t=1}^T$ are asymptotically the same as the unobservable innovations $\left\{ \epsilon_t \right\}_{t=1}^T$. 

Therefore, the OLS-CUSUM statistic based on the standardized residuals is expressed as below 
\begin{align}
\mathcal{C}_T(k) = \underset{ 0 \leq k \leq T }{ \mathsf{max} }  \left\{ \frac{ \displaystyle  \sum_{t=1}^{k} \widehat{\epsilon}_t - \frac{k}{T} \sum_{t=1}^T \widehat{\epsilon}_t }{ \displaystyle    \widehat{\sigma}_{\epsilon} \sqrt{T} }  \right\} 
\equiv 
\underset{ 0 \leq s \leq 1 }{ \mathsf{sup} } \left\{  \frac{1}{ \widehat{\sigma}_{\epsilon} } \left( \frac{1}{\sqrt{T} } \sum_{t=1}^{ \floor{Ts} } \widehat{\epsilon}_t  - \frac{s}{\sqrt{T}} \sum_{t=1}^T \widehat{\epsilon}_t  \right) \right\} 
\end{align}
Then, it can be shown that $\mathcal{C}_T(k) \Rightarrow \big[ W(s) - s W(1) \big]$ which is Brownian bridge. In particular, the weakly convergence of the OLS-CUSUM statistic is based on $\underset{ 0 \leq k \leq T }{ \mathsf{max} } \mathcal{C}_T(k) \Rightarrow \underset{ 0 \leq s \leq 1 }{ \mathsf{sup} } \big[ W(s) - s W(1) \big]$.

\medskip

\begin{remark}
Notice that Example \ref{Example1} considers a linear  time series model with no covariates, thus the structural break statistic can be considered as a difference in means for the two sub-samples, under the null hypothesis. The partial sum process for the innovation sequences of the model follows standard Brownian motion limit results.  Furthermore, in Example \ref{Example2} we consider the OLS-CUSUM statistic for the classical regression model as proposed by the seminal study of \cite{ploberger1992cusum}.
\end{remark}

%%-------------------------------------------------------------------------%%
\newpage 

\begin{example}
\label{Example2}
Consider the following time series model 
\begin{align}
y_t = x_t^{\prime} \beta_1 \mathbf{1} \{ t \leq k \}  + x_t^{\prime} \beta_2 \mathbf{1} \{ t > k \} + \epsilon_t, \ t =1,...,T  
\end{align}
where $y_t$ is the dependent variable, $x_t = \big[ 1, x_{2,t},...,x_{p,t} \big]^{\prime} = \big[ 1, \tilde{x}_t^{\prime} \big]^{\prime}$ is a $(p+1)-$dimensional vector and the error term $\epsilon_t$ are assumed to be  \textit{i.i.d} $(0, \sigma_\epsilon^2)$. We define $x_{1,t} \equiv   x_t^{\prime} \mathbf{1} \{ t \leq k \}$ and $x_{2,t} \equiv x_t^{\prime} \mathbf{1} \{ t > k \}$ for $k = [Ts]$ with $s \in [0,1]$. Under the null hypothesis of no structural break $\mathbb{H}_0: \beta_1 = \beta_2$ with $\widehat{\beta}_T$ the $\sqrt{T}$-consistent estimator for $\beta$ such that $\sqrt{T} \left( \hat{\beta}_T - \beta \right) = \mathcal{O}_p(1)$ is bounded in probability. 
\end{example}

Based on Example \ref{Example2} the OLS-CUSUM statistic, $\mathcal{C}_T(k)$, is obtained using the OLS residuals under the null hypothesis, defined as $\widehat{\epsilon}_t = y_t - \widehat{\beta}_T x_t = \epsilon_t - x_t^{\prime} \left( \widehat{\beta}_T - \beta  \right)$. Then, we obtain that 
\begin{align}
\mathcal{C}_T(k) = \frac{1}{ \widehat{\sigma}_{\epsilon} } \left(  \frac{1}{\sqrt{T} } \sum_{t=1}^{ \floor{Ts} } \widehat{\epsilon}_t - \frac{ \floor{Ts} }{T}  \frac{1}{\sqrt{T}} \sum_{t=1}^T \widehat{\epsilon}_t  \right)
\end{align}
In particular, we aim to show that $\mathcal{C}_T(k) {\Rightarrow} \big[ W(s) - s W(1) \big]$, which shows weakly convergence of the statistic to the Brownian bridge process. To do this, we consider the OLS residuals which can be expressed as below
\begin{align}
\label{OLSresiduals}
\frac{1}{ \sqrt{T} } \sum_{t=1}^{ \floor{Ts} }  \widehat{ \epsilon}_t = \frac{1}{ \sqrt{T} } \sum_{t=1}^{ \floor{Ts} } \epsilon_t - \frac{1}{ \sqrt{T} }  \sum_{t=1}^{ \floor{Ts} }  x_t^{\prime} \left( \widehat{\beta}_T - \beta  \right)
\end{align}
Furthermore, the following result holds  
\begin{align}\label{PM1992}
\frac{1}{\sqrt{T}} \sum_{t=1}^{ \floor{Ts} } x_t^{\prime} \left( \widehat{\beta}_T - \beta  \right) = \frac{r}{\sqrt{T}} \sum_{t=1}^T \epsilon_t + o_p(1).
\end{align}
A short proof on the asymptotic result above is provided here. We can express the left side of (\ref{PM1992}) as an inner product since our framework allows such representation
\begin{align}
\left[ \frac{1}{\sqrt{T}} \sum_{t=1}^{ \floor{Ts} } x_t^{\prime} \left( \widehat{\beta}_T - \beta  \right) \right] = \left[  \frac{1}{T} \sum_{t=1}^{ \floor{Ts} } x_t^{\prime} \right] . \left[ \sqrt{T} \left( \widehat{\beta}_T - \beta  \right) \right]
\end{align}
Notice that it holds that 
\begin{align}
\left(  \frac{1}{T} \sum_{t=1}^{ \floor{Ts} } x_t^{\prime} \right) \overset{p}{\to} [s, 0,...,0], \ \ \text{since} \ \ \underset{ T \to \infty }{ \text{lim}} \frac{1}{T} \sum_{t=1}^{ \floor{Ts} } \tilde{x}_t = 0 \ \ \text{and} \ \ \ \underset{ T \to \infty }{ \text{lim}} \frac{1}{T} \sum_{t=1}^{\floor{Ts} } 1 = s
\end{align}
Then, the second term with a matrix decomposition for $Q = \left( \frac{1}{T} \sum_{t=1}^T x_t x_t^{\prime} \right)$ where $x_t= \big[ 1, \tilde{x}_t^{\prime} \big]^{\prime}$ can be expressed as below
\begin{align}
\sqrt{T} \left( \widehat{\beta}_T - \beta  \right)  
= 
\left( \frac{1}{T} \sum_{t=1}^T x_t x_t^{\prime} \right)^{-1} \left( \frac{1}{\sqrt{T}} \sum_{t=1}^T x_t \epsilon_t  \right) 
\equiv
\frac{1}{ \sqrt{T} }
\begin{bmatrix}
1 & \boldsymbol{0} 
\\
\\
\boldsymbol{0} & \tilde{ \boldsymbol{Q} }
\end{bmatrix}^{-1} 
\begin{bmatrix}
\displaystyle \sum_{t=1}^T  \epsilon_t 
\\
\displaystyle \sum_{t=1}^T  \tilde{x}_t \epsilon_t
\end{bmatrix} + o_p(1)
\end{align}
since $\underset{ T \to \infty }{ \text{lim}} \frac{1}{T} \sum_{t=1}^{ \floor{Ts} } \tilde{x}_t \tilde{x}_t^{ \prime} = \tilde{\boldsymbol{Q}}$. Also, note that $\begin{bmatrix}
1 & \boldsymbol{0} 
\\
\boldsymbol{0}  & \tilde{ \boldsymbol{Q} }
\end{bmatrix}^{-1} = 
\begin{bmatrix}
1 & \boldsymbol{0} 
\\
\boldsymbol{0}  & \tilde{ \boldsymbol{Q} }^{-1}
\end{bmatrix}$.

%%-------------------------------------------------------------------------%%
\newpage 

Therefore, we obtain
\begin{align}
\left[ \frac{1}{\sqrt{T}} \sum_{t=1}^{ \floor{Ts} } x_t^{\prime} \left( \widehat{\beta}_T - \beta  \right) \right]  \overset{p}{\to} [s \ \boldsymbol{0} ] \frac{1}{ \sqrt{T} }  \begin{bmatrix}
1 & \boldsymbol{0} 
\\
\\
\boldsymbol{0} & \tilde{ \boldsymbol{Q} }^{-1}
\end{bmatrix} 
\begin{bmatrix}
\displaystyle  \sum_{t=1}^T  \epsilon_t 
\\
\displaystyle   \sum_{t=1}^T  \tilde{x}_t \epsilon_t
\end{bmatrix} 
= 
\frac{s}{\sqrt{T}} \sum_{t=1}^T \epsilon_t + o_p(1)
\end{align}
where $\boldsymbol{0}$ is $(p-1)$ dimensional column vectors of zeros. Then, using the asymptotic result given by (\ref{PM1992}) as well as the expression for the OLS residuals as in (\ref{OLSresiduals}) the OLS-CUSUM statistic can be expressed as
\begin{align*}
\mathcal{C}_T(k) 
&= 
\frac{1}{ \widehat{\sigma}_{\epsilon} } \frac{1}{ \sqrt{T} } \left\{ \left( \sum_{t=1}^{k} \epsilon_t -  \sum_{t=1}^{k} x_t^{\prime} \left( \widehat{\beta}_T - \beta  \right) \right) - s \left( \sum_{t=1}^{T} \epsilon_t - \sum_{t=1}^{T} x_t^{\prime} \left( \widehat{\beta}_T - \beta \right) \right) \right\} 
\\
&= \frac{1}{ \widehat{\sigma}_{\epsilon} } \frac{1}{ \sqrt{T} } \left\{  \left( \sum_{t=1}^{k} \epsilon_t -        s \sum_{t=1}^{T} \epsilon_t \right) - \left(  \sum_{t=1}^{k} x_t^{\prime} \left( \widehat{\beta}_T - \beta  \right) - s \sum_{t=1}^{T} x_t^{\prime} \left( \widehat{\beta}_T - \beta  \right) \right) \right\}
\end{align*}
Notice that the second term above converges to zero in probability such that 
\begin{align}
s \left( \sum_{t=1}^T \epsilon_t - \sum_{t=1}^T \widehat{\epsilon}_t \right) = o_p(1)
\end{align}
then the limit result follows as given below 
\begin{align}
\mathcal{C}_T(k) = \underset{ s \in [0, 1]  }{ \mathsf{sup} }  \left\{ \frac{1}{\widehat{\sigma}_{\epsilon} } \left(   \sum_{t=1}^{k} \frac{ \epsilon_t }{ \sqrt{T} } - s \sum_{t=1}^{T} \frac{ \epsilon_t }{  \sqrt{T} } \right) \right\} \Rightarrow \big[ W(s) - s W(1) \big], \ \  s \in [0,1].
\end{align} 
Thus, the OLS-CUSUM statistic weakly converges to the Brownian bridge limit uniformly for $s \in [0,1]$.

\medskip

\begin{example}
Similarly to Examples \ref{Example1} and \ref{Example2} we can derive limit theory results for the OLS-CUSUM squared test (see detailed proofs in \cite{deng2008limit}). Following Definition \ref{definition1} and Theorem \ref{Theorem1} the corresponding test statistic is defined as below 
\begin{align}
\mathcal{C}^{(2)}_T(k) 
= 
\underset{ 0 \leq k \leq T }{ \mathsf{max} }  \left(  \frac{ \displaystyle  \sum_{t=1}^{k} \widehat{\epsilon}^2_t - \frac{k}{T} \sum_{t=1}^T \widehat{\epsilon}^2_t }{  \widehat{\sigma}_{\epsilon} \sqrt{T} }  \right)
=  
\underset{ 0 \leq s \leq 1 }{ \mathsf{sup} }  \left\{  \frac{1}{ \widehat{\sigma}_{\epsilon} } \left(   \frac{1}{\sqrt{T}} \sum_{t=1}^{ \floor{Ts} } \widehat{\epsilon}^2_t   -  \frac{s}{\sqrt{T}} \sum_{t=1}^T \hat{\epsilon}^2_t  \right) \right\} 
\end{align}
Furthermore, the following two sufficient conditions hold
\begin{align}
\underset{ 1 \leq j \leq n  }{ \mathsf{max} } \left|  \frac{1}{ \sqrt{T} } \sum_{t=1}^j \epsilon_t x_t^{\prime} \left( \widehat{\beta}_T - \beta \right) \right| \overset{ p }{ \to } 0 \\
\underset{ 1 \leq j \leq n  }{ \mathsf{max} } \ \frac{1}{ \sqrt{T} } \left( \widehat{\beta}_T - \beta \right)^{\prime} \sum_{t=1}^j x_t x_t^{\prime}  \left( \widehat{\beta}_T - \beta \right) \overset{ p }{ \to } 0
\end{align}
Therefore, it can be proved that the weakly convergence result to a Brownian bridge follows 
\begin{align}
\mathcal{C}^{(2)}_T(k) \Rightarrow \underset{ 0 \leq s \leq T }{ \mathsf{sup} } \ \big[ W(s) - s W(1) \big].
\end{align}
\end{example}

%%-------------------------------------------------------------------------%%
\newpage 

\section{WALD TYPE STATISTICS}
\label{Section3}

In this section we examine the implementation of Wald-type statistics when constructing an equivalent structural break test for the conditional mean of a regression model. The particular methodology has been advanced by the seminal work of \cite{hawkins1987test} and \cite{andrews1993tests}. 

\begin{definition}
\label{definition2}
We define the following $p-$dimensional limiting distribution for some $\pi \in (0,1)$ such that 
\begin{align}
\mathcal{Q}_p( \pi ) 
:= 
\frac{ \bigg[ W_p(\pi) - \pi W_p(1) \bigg]^{\prime} \bigg[ W_p(\pi) - \pi W_p(1) \bigg] }{ \pi ( 1 - \pi ) } 
\end{align}
where $W_p(.)$ is $p \times 1$ vector of independent standard Brownian motions for some $p \geq 1$, where $p$ represents the number of parameters that the null hypothesis is testing for stability. 
\end{definition}

\medskip

\begin{remark}
The limit process $\mathcal{Q}_p( \pi )$ is referred to as the square of a standardized tied-down Bessel process of order $p$ (see, \cite{andrews1993tests}). In particular, the limit process given by Definition \ref{definition2} a is employed when deriving the asymptotic distribution of sup-Wald statistics, such that $\underset{ \pi \in (0,1) }{ \mathsf{sup} } W_T( \pi ) \overset{ d }{ \to } \underset{ \pi \in (0,1) }{ \mathsf{sup} } \mathcal{Q}_p( \pi )$.  
\end{remark}

We mainly consider univariate regression models but the framework can be also generalized to time series regressions with multiple regressors. Furthermore, when applying the supremum functional to Wald-type statistics we assume that the unknown break-point $\pi \in (0,1)$ lies in a symmetric subset of the particular unit set, to ensure that inference is not at the boundary of the parameter space (see,  \cite{andrews2001testing}). 

\subsection{Unconditional Mean test}

Consider the following time series regression model 
\begin{align}
y_t 
= 
\theta_1 \mathbf{1} \{ t \leq k \}  + \theta_2 \mathbf{1} \{ t > k \} + \epsilon_t, \ t = 1,...,T  
\end{align}
where $\theta_1$ and $\theta_2$ are both deterministic and the break point $k = \floor{Ts}$ for some $s \in (0,1)$ is an unknown fixed fraction of the full sample period. The following assumption hold: 

\begin{assumption}\label{Assumption1}
Let $\epsilon_t$ be a sequence of random variables. Then, the following moment conditions hold. 
\begin{enumerate}
\item[A1.] $\mathbb{E} \left( \epsilon_t | \mathcal{F}_{t-1} \right) = 0$ with an asymptotic variance as $T \to \infty$ given as below
\begin{align}
\mathsf{var}\left( T^{-1/2} \sum_{t=1}^{ \floor{Ts} } \epsilon_t \right) 
= 
\frac{ \floor{Ts}  }{T} \sigma^2_{\epsilon} \overset{ p }{ \to } 
s \sigma^2_{\epsilon}, \ \ \ \sigma^2_{\epsilon} = \underset{ T \to \infty }{ \mathsf{lim} } \frac{1}{T} \mathbb{E} \left[ \left( \sum_{t=1}^T \epsilon_t \right)^2 \right]
\end{align}

\item[A2.] sup$_t$ $\mathbb{E} \left( |  \epsilon_t | \right)^{2 + d} < \infty$ for some $d > 2$.

\item[A3.] $\{ \epsilon_t \}_{t=1}^T$ satisfies the following \textit{Functional Central Limit Theorem (FCLT}),
\begin{align}
\frac{1}{\sqrt{T}} \sum_{t=1}^{ \floor{Ts} } \epsilon_t \overset{\mathcal{D}}{\to} \sigma_{\epsilon} W(s) \equiv B(s)
\end{align} 
\end{enumerate}
\end{assumption}

%%-------------------------------------------------------------------------%%
\newpage 

\begin{remark}
Assumption A1 gives the moment conditions of the classical regression model with only intercept. In particular, $\mathsf{Var} \left( T^{-1/2} \sum_{t=1}^{ \floor{Ts} } \epsilon_t \right)$ gives the asymptotic variance of the partial-sum of innovations. Since the limiting variance is bounded then convergence in distribution follows.    
\end{remark}
The null and alternative hypothesis are as below
\begin{align}
\mathbb{H}_0: \theta_1 = \theta_2 \ \text{versus} \ \mathbb{H}_A: \theta_1 \neq \theta_2 
\end{align}
and we consider the following test statistic for detecting structural change in the unconditional mean of the simple linear regression with only intercept. 
\begin{align}\label{Zn}
\mathcal{Z}_T (s) 
&= 
\underset{ 0 \leq k \leq T }{ \mathsf{max} }  \left\{  T \frac{ \left( \bar{y_{1}} - \bar{y_{2}} \right)^2 }{ \hat{\sigma}^2_{ z } } \right\} 
\equiv 
\underset{ s \in ( 0 , 1 )  }{ \mathsf{sup} } \left\{  T \frac{ \left( \bar{y_{1}} - \bar{y_{2}} \right)^2 }{ \widehat{\sigma}^2_{ z } } \right\} + o_p(1) 
\\
\bar{y_{1}} 
&= \frac{1}{ k } \sum_{ t = 1 }^{ k } y_t \ \ \ \text{and} \ \ \ \bar{y_{2}}
= 
\frac{1}{ T - k } \sum_{ t = k + 1 }^{ T } y_t  \ \  \text{where} \ \ \widehat{\sigma}^2_{ z }
= 
\frac{ \widehat{\sigma}^2_{\epsilon} }{ \frac{k}{T} \left( 1 - \frac{k}{T} \right) } \overset{p}{\to} \frac{ \sigma^2_{\epsilon} }{ s (1-s) }
\end{align}
We can show that the asymptotic variance of the normalized statistical distance measure $\sqrt{T} \left( \bar{y_{1}} -  \bar{y_{2}} \right)$  is given  by  $\mathsf{Avar} \left[\sqrt{T} \left( \bar{y_{1}} -  \bar{y_{2}} \right) \right] = \frac{ \sigma^2_{\epsilon} }{ s (1-s) } \equiv \sigma^2_{z}$ by noting that $\underset{ T \to \infty }{ \text{lim} }  \frac{ T }{ \floor{Ts} } = \frac{1}{s}$ and $\underset{ T  \to \infty }{ \text{lim} }  \frac{ T }{ T - \floor{Ts}} = \frac{1}{s(1-s)}$. According to \cite{aue2013structural} testing the null hypothesis of equal means across a $p$-dimensional multivariate time series is equivalent to constructing a $p-$dimensional CUSUM process and testing for a structural break at an unknown break point $k$. Due to the fact that the partial-sum process representing the CUSUM statistic has a weak convergence to a $p-$dimensional Brownian Motion process, then the quadratic form of the partial-sum process weakly convergence to the sum of squared independent Brownian bridges. The specific property permits to establish an equivalence between a supremum Wald-type statistic and a CUSUM-type statistic. In particular, we show that the asymptotic distribution of the OLS-CUSUM statistic can be deduced from the asymptotic distribution of $\mathcal{Z}_T$ and vice-versa.  

\begin{theorem}
\label{theorem2}
Let $\mathcal{Z}_T(s)$ be the test statistic for testing the null hypothesis $\mathbb{H}_0: \theta_1 = \theta_2$ of no structural change in the unconditional mean as defined by (\ref{Zn}) for some $0 \leq k \leq T$. Assume the conditions A1 to A3 given by Assumption \ref{Assumption1} hold and let $\mathcal{Z}^o_T(s)$ be an OLS-CUSUM type statistic. Then, both statistics weakly converge in the Skorokhod space $\mathcal{D}[0,1]$ to functionals of a Brownian bridge such as $\{ \mathcal{BB}(s): s \in [0,1] \}$ where $\mathcal{BB}(s) =  \big[ W(s) - s W(1) \big]$.
\begin{align}
\mathcal{Z}_T (r) 
&= 
\underset{ s \in [\nu , 1 - \nu]  }{ \mathsf{sup} } \left\{ T \frac{ \left( \bar{y_{1}} - \bar{y_{2}} \right)^2 }{ \widehat{\sigma}^2_{ z } } \right\}  \Rightarrow \underset{ s \in [\nu , 1 - \nu]  }{ \mathsf{sup} } \frac{ \big[ W(s) - s W(1) \big]^2 }{s(1-s)} 
\\
\mathcal{Z}^o_T(s) 
&=  
\underset{ s \in [\nu , 1 - \nu]  }{ \mathsf{sup} } \left\{  \frac{1}{ \sqrt{T} } \frac{1}{\sqrt{ \widehat{\sigma}_{\epsilon}  } } \left( \sum_{t=1}^{ \floor{Ts} } y_t - s \sum_{t=1}^T y_t  \right) \right\} \Rightarrow \underset{ s \in [\nu , 1 - \nu]  }{ \mathsf{sup} } \big[ W(s) - s W(1) \big] 
\end{align}
\end{theorem}

\begin{remark}
Theorem 2 implies that the $\mathcal{Z}_T(s)$ statistic of equality of means in a SLR model with only intercept across the unknown break point has an equivalent weak convergence to the CUSUM statistic of the time series under consideration using a suitable normalization constant. 
\end{remark}

Next, we consider the corresponding supremum Wald-type statistic for testing for structural change in the unconditional mean of the classical regression model with only intercept. The test is formulated as below
\begin{align}
\mathcal{W}_T 
= 
\frac{1}{ \widehat{\sigma}_{\epsilon}^2 } \left( \mathcal{ \boldsymbol{R} } \boldsymbol{\Theta} \right)^{ \prime } \left[ \mathcal{ \boldsymbol{R} } \left( \boldsymbol{Z}^{\prime} \boldsymbol{Z}  \right)^{-1}  \mathcal{\boldsymbol{R} }^{\prime}  \right]^{-1} \left( \mathcal{ \boldsymbol{R} } \boldsymbol{\Theta} \right)
\end{align}

%%-------------------------------------------------------------------------%%
\newpage 

where $\widehat{\sigma}_{\epsilon}^2 = \frac{1}{T} \sum_{t=1}^T \hat{\epsilon}_t^2(k) $ the residual variance under the null hypothesis and is a consistent estimator of $\sigma_{\epsilon}$ such that $\widehat{\sigma}_{\epsilon}^2 \overset{ p }{ \to } \sigma_{\epsilon}^2$. Thus, by substituting the restriction matrix $\mathcal{\boldsymbol{R}} = \big[ \boldsymbol{I} - \boldsymbol{I} \big]$, $\boldsymbol{Z} = \big[ \boldsymbol{X}_1 \ \boldsymbol{X}_2 \big]^{\prime}$ and the estimator $\widehat{\boldsymbol{\Theta} } =  \big[ \widehat{\theta}_1 \ \widehat{\theta}_2 \big]^{\prime}$ of $\boldsymbol{\Theta}$ into the above formulation of the Wald statistic we obtain the expression 
\begin{align}
\mathcal{W}_T( s )
&= 
\frac{1}{\widehat{\sigma}^2_{\epsilon} } \left( \widehat{\theta}_1 - \widehat{\theta}_1 \right)^{\prime} \left[ \left( X_1^{\prime} X_1 \right)^{-1} +  \left( X_2^{\prime} X_2 \right)^{-1} \right]^{-1} \left( \widehat{\theta}_1 - \widehat{\theta}_1 \right)
\end{align}
We observe that for the unconditional mean model it holds that  
\begin{align*}
\left( X_1^{\prime} X_1 \right) = \sum_{t=1}^T \mathbf{1} \{ t \leq k \} = k \ \ \text{and} \ \ \left( X_2^{\prime} X_2 \right) = \sum_{t=1}^T \mathbf{1} \{ t > k \} = T - k
\end{align*} 
Since in this section we consider mean shifts, $X_1$ stacks the elements of $x_t$ for which $t \leq k$, that is, $\mathbf{1} \{ t \leq k \}$ and $X_2$ stacks the elements of $x_t$ for which $t > k$, that is, $\mathbf{1} \{ t \leq k \}$. Therefore, using the corresponding matrix notation the OLS estimators can be written as below
\begin{align*}
\widehat{\theta}_1 = \left( X_1^{\prime} X_1 \right)^{-1}  X_1^{\prime} y = \frac{1}{k} \sum_{t=1}^k y_t  , \ \ \ 
\widehat{\theta}_2 = \left( X_2^{\prime} X_2 \right)^{-1}  X_2^{\prime} y = \frac{1}{T-k} \sum_{t=1}^T y_t 
\end{align*}
Thus, the Wald statistic can be expressed as 
\begin{align*}
\mathcal{W}_T (s ) = \frac{1}{ \widehat{\sigma}^2_{\epsilon} } \left( \bar{y_{1}} -  \bar{y_{2}} \right)^2 \left[ \frac{1}{k} + \frac{1}{T-k} \right]^{-1} = \frac{k(T-k) }{ T \widehat{\sigma}_{\epsilon}^2 } \left( \bar{y_{1}} - \bar{y_{2}} \right)^2
\end{align*}
Under the null hypothesis $\mathbb{H}_0: \theta_1 = \theta_2$, we have that since $\left( \bar{y_{1}} -  \bar{y_{2}} \right) \equiv \left( \widehat{\theta}_1 - \widehat{\theta}_2 \right)$ we obtain the expression 
\begin{align*}
\left( \widehat{\theta}_1 - \widehat{\theta}_2 \right) = \frac{ T }{ k(T-k) } \left( \sum_{t=1}^k \epsilon_t - \frac{k}{T}  \sum_{t=1}^T \epsilon_t \right) = \frac{ T \sqrt{T} }{ k(T-k) } \left(  \sum_{t=1}^k \frac{\epsilon_t}{\sqrt{T}}   - \frac{k}{T}  \sum_{t=1}^T \frac{\epsilon_t}{\sqrt{T}} \right)
\end{align*}
Therefore, the Wald test is equivalently written as below
\begin{align*}
\mathcal{W}_T (s ) 
&= 
\frac{k(T-k) }{ T \widehat{\sigma}^2_{\epsilon} } \left\{  \frac{T  \sqrt{T} }{ k(T-k) } \left(  \sum_{t=1}^k \frac{\epsilon_t}{\sqrt{T}}   - \frac{k}{T}  \sum_{t=1}^T \frac{\epsilon_t}{\sqrt{nT}} \right) \right\}^2 
\\
&= 
\frac{ T }{ \frac{k}{T} \left( 1 - \frac{k}{T} \right) } \frac{1}{ \widehat{\sigma}^2_{\epsilon} } \left(  \sum_{t=1}^k \frac{\epsilon_t}{\sqrt{T}} - \frac{k}{T} \sum_{t=1}^T \frac{\epsilon_t}{\sqrt{T}} \right)^2.
\end{align*} 
A simple application of the WLLN for the variance of the OLS estimator implies that $\widehat{\sigma}^2_{\epsilon} \overset{ p }{ \to } \sigma_{\epsilon}$ as $T \to \infty$. Moreover since it holds that $\frac{k}{T} \frac{T-k}{T} \overset{ p }{ \to } s(1-s)$, we obtain the weak convergence for the sup-Wald statistic
\begin{align}
\mathcal{W}_T^{*} (s) 
:= 
\underset{ s \in [\nu , 1 - \nu]  }{ \mathsf{sup} } \mathcal{W}_T (s ) \Rightarrow \underset{ s \in [\nu , 1 - \nu]  }{ \mathsf{sup} } \frac{ \left[ W(s) - s W(1) \right]^{\prime} \left[ W(s) - s W(1) \right] }{s(1-s)}   .
\end{align}
The above asymptotic result indeed verifies the weakly convergence of the sup-Wald statistic into the supremum of  a normalized squared Brownian bridge, specifically for the unconditional mean specification of the regression model. Then, sstatistical inference is conducted based on the null hypothesis, $\mathbb{H}_0: \theta_1 = \theta_2$, which is rejected for large values of the sup-Wald statistic with a significance level $\alpha \in (0,1)$. Thus, the exact form of the limiting distribution of the statistic is employed to obtain associated critical values, denoted with $c_{\alpha}$ such that $\mathbb{P} \left( \mathcal{W}_T^{*} (s) > c_{\alpha} \right) > 0$ with $\underset{ T \to \infty }{ \mathsf{lim} } \mathbb{P} \left( \mathcal{W}_T^{*} (s) > c_{\alpha} \right) = 1$.

%%-------------------------------------------------------------------------%%
\newpage

\subsection{Conditional Mean test}

Consider the following model 
\begin{align}
\label{Conditional}
y_t = x_t^{\prime} \theta_1 \mathbf{1} \{ t \leq k  \}  + x_t^{\prime} \theta_2 \mathbf{1} \{ t > k \} + \epsilon_t, \ \ \ t = 1,...,T  
\end{align}
where $y_t$ is the dependent variable, $x_t$ is a $p \times 1$ vector of regressors, $\epsilon_t$ is an unobservable disturbance term with $\mathbb{E}\left( \epsilon_t | x_t \right) = 0$ almost surely and $\theta_1$ and $\theta_2$ the regression coefficients formulated under the null hypothesis of no structural break. Define with $x_{1,t} \equiv   x_t^{\prime} \mathbf{1} \{ t \leq k \}$ and $x_{2,t} \equiv   x_t^{\prime} \mathbf{1} \{ t > k \}$ where $k = \floor{Ts}$ for some $s \in (0,1)$. Furthermore, notice that the regressor vector $x_t$ can contain exogenous regressors and lagged dependent variables with unknown integration order. Moreover, under suitable regularity assumptions one can consider a static, a dynamic time series regression model as well as models with integrated regressors or cointegrated regressors. 

The null hypothesis of interest is 
\begin{align}
\mathbb{H}_0: \theta_1 = \theta_2 \equiv \theta_0 \in \mathbb{R}^{p \times 1} \ \text{for all} \ t. 
\end{align}
The alternative hypothesis $\mathbb{H}_A$ is that $\mathbb{H}_0$ is false, which implies that the regression coefficient has a structural break at some unknown break point in the sample. Under $\mathbb{H}_0$, the unknown constant parameter vector $\theta$ can be consistently estimated using the OLS estimator such that
\begin{align}
\widehat{\theta} 
= 
\underset{ \theta \in \mathbb{R}^{p \times 1} }{\mathsf{arg \ min } } \sum_{t=1}^T \left( y_t - x_t^{\prime} \theta_0 \right)^2 
\end{align} 
Under the alternative $\mathbb{H}_A$, $\theta_0 \equiv \theta_t$ can be considered as a time-varying parameter vector, which implies, $\theta_t \equiv \theta_1$ for $1 \leq  t \leq k$ and $\theta_t \equiv \theta_2$ for $k + 1 \leq  t \leq T$. To facilitate the estimation and inference based on the above regression model we impose the following regularity conditions. 

\begin{assumption}
\label{Assumption1}
Let $\epsilon_t$ be a sequence of random variables. Then the following moment conditions and FCLT  hold, which allow $x_t$ and $\epsilon_t$ to be weakly dependent. 
\begin{enumerate}

\item[B1.] $\{\epsilon_t\}_{t=1}^{T}$ is a homoskedastic \textit{martingale difference sequence (m.d.s)}  such that $\mathbb{E} \left( \epsilon_t | \mathcal{F}_{t-1} \right) = 0$ and $\mathbb{E}\left( \epsilon_t^2  \right) =\sigma^2$, where $\mathcal{F}_{t-1} = \left\{ x^{\prime}_t, x^{\prime}_{t-1},..., \epsilon_{t-1}, \epsilon_{t-2},... \right\}$.

\item[B2.] $\{x_t\}_{t=1}^{T}$ has at least a finite second moment, that is, sup$_t$ $\sum_{t=1}^{T} ||x_t||^{2 + d} < \infty$ where $d > 0$ and the following \textit{weak law of large numbers (WLLN)} holds
\begin{align}
\frac{1}{T} \sum_{t=1}^{T} x_t x_t^{\prime} \overset{p}{\to} \boldsymbol{Q}_T \ \ \text{as} \ \ T \to \infty  \ \text{such that } \ \underset{ s \in  [0,1] }{ \mathsf{sup} } \bigg\rvert  \frac{1}{T} \sum_{t=1}^{ \floor{Ts}  } x_t x_t^{\prime} - s\boldsymbol{Q}_T    \bigg\rvert = o_p(1) 
\end{align}
where $\boldsymbol{Q}_T$ is a $(p \times p)$ non-stochastic finite and positive definite matrix.

\item[B3.] $\{x_t \epsilon_t \}_{t=1}^{T}$ satisfies the following \textit{Functional Central Limit Theorem\footnote{The multivariate FCLT is examined in the studies of \cite{Phillips1986multiple} and  \cite{wooldridge1988some}.}(FCLT)}, 
\begin{align}
\mathcal{S}_T(s) = \left( T^{-1/2} \boldsymbol{\Omega}_T^{-1/2} \sum_{t=1}^{ \floor{Ts} } x_t \epsilon_t \right) \Rightarrow  \boldsymbol{W}_p(s), \ s \in (0,1) \ \text{and} \ \boldsymbol{\Omega}_T = \mathbb{E} \left( x_t x_t^{\prime} \epsilon^2_t  \right) >0.
\end{align}
\end{enumerate}
\end{assumption}

%%-------------------------------------------------------------------------%%
\newpage 

In other words, the general moment conditions given by Assumption \ref{Assumption1} permits to consider both residual-based statistics as well as Wald-type statistics as suitable detectors when testing for a break-point within the full sample. In particular, \cite{chen2012testing} consider the implementation of generalized Hausman-type tests using nonparametric estimation and inference techniques.

\subsubsection{Asymptotic Convergence of sup-Wald test}

Next, we derive the asymptotic convergence of the Wald test when detecting a single structural break in the regression for the classical regression model with a conditional mean specification form, under Assumptions B1 to B3. Furthermore, under the null hypothesis the break-point is unidentified, thus to facilitate statistical inference we consider the corresponding supremum Wald-type statistic based on the OLS estimator which is expressed as $\mathcal{W}^{*}_T(s) := \underset{ s \in [ \nu, 1 - \nu] }{ \text{sup} } \mathcal{W}_T(s)$ for some $\nu \in (0,1)$. 

\medskip

Using the null hypothesis $\mathbb{H}_0: \theta_1 = \theta_2$ of no structural break we obtain an equivalent representation using the linear restriction matrix $\mathcal{ \boldsymbol{R} }$ of rank $q$, which implies that $\mathbb{H}_0: \mathcal{\boldsymbol{R} } \boldsymbol{\Theta} = \boldsymbol{0}$. We define $\boldsymbol{\Theta} = \big[ \theta_1 \ \theta_2 \big]^{\prime}$, $\boldsymbol{Z} = \big[ \boldsymbol{X}_1 \ \boldsymbol{X}_2 \big]^{\prime}$ and $\mathcal{\boldsymbol{R}} = \big[ \boldsymbol{I} - \boldsymbol{I}  \big]$ and prove that the asymptotic distribution of the corresponding sup-Wald statistic weakly convergences to the supremum of a normalized squared Brownian bridge. 

\begin{theorem}
\label{theorem3}
Under the null hypothesis $\mathbb{H}_0: \theta_1 = \theta_2$ with $0 < \nu < 1$ we define $\mathcal{W}^{*}_T(s)$
\begin{align}
\mathcal{W}^{*}_T(s) := \underset{ s \in [ \nu, 1 - \nu] }{ \text{sup} } \mathcal{W}_T(s) \Rightarrow   \underset{ s \in [\nu , 1 - \nu]  }{ \mathsf{sup} } \frac{  \big[ \boldsymbol{W}_p(s) - s \boldsymbol{W}_p(1) \big]^{\prime} \boldsymbol{Q}_T^{-1}  \big[ \boldsymbol{W}_p(s) - s \boldsymbol{W}_p(1) \big] }{s(1-s)} .
\end{align} 
where $\mathcal{W}_T(s)$ denotes the Wald statistic for testing the null hypothesis $\mathbb{H}_0: \theta_1 = \theta_2$ and is expressed as
\begin{align}
\mathcal{W}_T 
= 
\frac{1}{ \widehat{\sigma}_{\epsilon}^2 } \left( \mathcal{ \boldsymbol{R} } \boldsymbol{\Theta} \right)^{ \prime } \left[ \mathcal{ \boldsymbol{R} } \left( \boldsymbol{Z}^{\prime} \boldsymbol{Z}  \right)^{-1}  \mathcal{\boldsymbol{R} }^{\prime}  \right]^{-1} \left( \mathcal{ \boldsymbol{R} } \boldsymbol{\Theta} \right)
\end{align}
with $\widehat{\sigma}^2_{\epsilon} = \displaystyle \frac{1}{T} \sum_{t=1}^T \left(  y_t - x_t^{\prime} \widehat{\theta}_T \right)$ a consistent estimate of $\sigma^2_{\epsilon}$ the OLS variance under the null hypothesis. 
\end{theorem}

\medskip

\begin{remark}
The above asymptotic theory analysis demonstrates that CUSUM-type statistics for detecting a structural break in the classical regression model with a single unknown break-point typically weakly converge to Brownian bridge functionals, such that $\mathcal{BB}(s) := \big[ W_p(s) - sW_p(1) \big]$ while the corresponding Wald-type tests weakly converge to a normalized version of the CUSUM test with a normalized constant given by $k_T := \left\{ \frac{k}{T} \left( 1 - \frac{k}{T} \right) \right\}^{1/2}$ where $k = \floor{Ts}$ for some $s \in [0,1]$. 
\end{remark}

Monte Carlo simulation experiments can be used to compare the asymptotic validity and performance of OLS-CUSUM type statistics and Wald-type tests by obtaining associated empirical size and power results. An important criticism of CUSUM-type statistics is that these tests are based on residuals under the null hypothesis which implies that the test is not designed with a specific alternative under consideration. Therefore, although the use of these tests can lead to a monotonically increasing power, in practise it can be stochastically dominated by the power of a Wald type test (see, also \cite{andreou2008restoring}). On the other hand the advantage of using a Wald-type statistic when testing for a structural break, is that the construction of the test allows to incorporate residuals obtained either under the null or under the alternative hypothesis, providing this way superior power performance. Intuitively, using the residual variance of the unrestricted model leads to better finite-sample power since the Wald-type statistic contains information from the alternative model (alternative hypothesis).

%%-------------------------------------------------------------------------%%
\newpage 

\section{STRUCTURAL BREAK TESTING UNDER NONSTATIONARITY}
\label{Section4}

Although the purpose of this paper is to explain the main challenges one faces when testing for a structural break in nonstandard econometric problems we discuss the main intuition when developing associated asymptotic theory using some examples. More specifically, in order to accommodate the nonstationary aspect in time series models\footnote{Standard regularity conditions for estimation and inference in nonstationarity time series models, is the asymptotic theory developed for nearly unstable autoregressive processes. The limit theory of time series models such as the asymptotic inference of AR(1) processes first examined by \cite{mann1943statistical} has been extended to the non-stationary asymptotic case as it captured via the local to unity framework which allows for the autoregressive coefficient of a univariate AR(1) to be expressed in the form $\rho = 1 + c/n$, where $n$ the sample size and $c$ the unknown degree of persistence of the data generating process.} we need to consider a suitable probability space in which weakly convergence arguments holds. Due to the fact that the asymptotic terms of sample moments of nonstationary time series models often involve non-standard limit results such as convergence to stochastic integrals, care is needed when applying related convergence arguments. In particular, the stochastic integrals of the form $\int_0^1 W dW (s)$ can be shown to converge weakly, under the null hypothesis, to the associated stochastic integral with the limiting Brownian motions, under the assumption of the existence of cadlag functionals in the unit interval equipped with the Skorokhod topology (see, \cite{muller2011efficient}).

\subsection{Linear Restrictions Testing}

We begin our analysis by discussing the main limit theory which is employed in nonstationary time series models, by providing two examples: \textit{(i)} a time series regression with integrated regressors and \textit{(ii)} a predictive regression with persistent regressors. Although, such asymptotics are commonly used when considering the asymptotic behaviour of t-tests and Wald-type tests based on linear restrictions on the parameter coefficients, these are applicable when constructing the corresponding structural break statistics.

\subsubsection{Time Series Regression with Integrated Regressors}

A class of nonstationary time series models include the linear regressions with integrated regressors as proposed by the studies of \cite{cavanagh1995inference} and \cite{jansson2006optimal} among others. 

\begin{example} (Cointegrating Regression,  \cite{banerjee1993co})

Consider the following bivariate system of co-integrated variables $\{ y_t \}_{t=1}^{\infty}$ and $\{ x_t \}_{t=1}^{\infty}$ such that  
\begin{align}
y_t &= \beta x_t + u_t \\
x_t &= x_{t-1} + \epsilon_t
\end{align}
with $u_t \sim N(0, \sigma^2_u )$, \ $\epsilon_t \sim N(0, \sigma^2_{\epsilon} )$ and $\mathbb{E}( u_t \epsilon_s ) = \sigma_{u \epsilon } \ \forall \ t \neq s$.
\end{example}
The OLS estimator of $\beta$ is given by $
\widehat{ \beta } = \left( \sum_{t=1}^T x_t^2 \right)^{-1} \left( \sum_{t=1}^T y_t x_t \right)$ which implies 
\begin{align}
T \left( \widehat{ \beta } - \beta \right)   = \left( T^{-2} \sum_{t=1}^T x_t^2 \right)^{-1} \left( T^{-1} \sum_{t=1}^T x_t u_t \right)
\end{align}
Thus, for this regression model due to the presence of the integrated regressor the limit is expressed as
\begin{align}
\left( \frac{1}{T^2} \sum_{t=1}^T x_t^2 \right) \Rightarrow \sigma^2_{ \epsilon } \int_0^1 W_{ \epsilon } (r)^2 dr.
\end{align}

%%-------------------------------------------------------------------------%%
\newpage 

Next, in order to derive the limiting distribution of the model parameter for the case of integrated regressors, we first consider the limiting distribution of the sample moment $\left( T^{-1} \sum_{t=1}^T x_t u_t \right)$. To do this, we assume the existence of a conditional distribution for $u_t$ given $\epsilon_t$, which implies the  conditional mean form
\begin{align}
u_t = \phi \epsilon_t + v_t, \ \ \phi = \frac{ \sigma_{u \epsilon} }{ \sigma^2_{\epsilon} } \ \ \text{and} \ \ \sigma^2_v = \sigma^2_u -   \frac{ \sigma^2_{u \epsilon} }{ \sigma^2_{\epsilon} } 
\end{align} 
Furthermore, we define with $W_{\epsilon} (r)$ and $W_{v} (r)$ to be two independent Wiener processes on $\mathcal{C}[0,1]$. Therefore, 
\begin{align}
\left( T^{-1} \sum_{t=1}^T x_t u_t \right)
= 
T^{-1}  \sum_{t=1}^T x_t \left(  \phi \epsilon_t + v_t \right) 
= 
\phi \left( T^{-1} \sum_{t=1}^T x_t \epsilon_t \right) + \left( T^{-1} \sum_{t=1}^T x_t v_t \right)
\end{align}
Substituting $x_t = x_{t-1} +  \epsilon_t$ into the above expression we obtain 
\begin{align*}
\left( T^{-1} \sum_{t=1}^T x_t u_t \right)
&= \phi \left( T^{-1} \sum_{t=1}^T (  x_{t-1} +  \epsilon_t ) \epsilon_t \right) + \left( T^{-1} \sum_{t=1}^T (  x_{t-1} +  \epsilon_t ) v_t \right)
\\
&= \phi \left( T^{-1} \sum_{t=1}^T x_{t-1} \epsilon_t \right) + \phi \left( T^{-1} \sum_{t=1}^T  \epsilon^2_t \right) 
\\
&\ \ \  + \left( T^{-1} \sum_{t=1}^T x_{t-1} v_t \right) + \left( T^{-1} \sum_{t=1}^T \epsilon_t v_t \right)
\end{align*}
Most importantly, within this setting the following asymptotic results hold
\begin{align}
T^{-1} \sum_{t=1}^T  \epsilon^2_t  & \overset{ p }{ \to } \sigma^2_{\epsilon}, \ \ \ \ \ \  T^{-1} \sum_{t=1}^T \epsilon_t v_t \overset{ p }{ \to }  0
\end{align}
and it has been also proved in the seminal study of \cite{Phillips1987time} (see, also  \cite{Phillips1986multiple}) 
\begin{align}
T^{-1} \sum_{t=1}^T x_{t-1} v_t & \Rightarrow \sigma_{ \epsilon } \sigma_{ v } \int_0^1 W_{\epsilon} (r) dW_v (r) \equiv \int_0^1 B_{\epsilon} (r) dB_v (r)
\\
T^{-1} \sum_{t=1}^T x_{t-1} \epsilon_t & \Rightarrow  \frac{ \sigma^2_{ \epsilon } }{2}  \bigg[  W^2_{\epsilon} (1) - 1 \bigg] 
\end{align}
Putting the above together we obtain that 
\begin{align}
T^{-1} \sum_{t=1}^T x_{t} u_t \Rightarrow \left\{ \phi \left(  \frac{ \sigma^2_{ \epsilon } }{2}  \bigg[  W^2_{\epsilon} (1) - 1 \bigg]  \right)   + \phi \sigma^2_{\epsilon} +   \sigma_{ \epsilon } \sigma_{ v } \int_0^1 W_{\epsilon} (r) dW_v (r)  \right\}
\end{align}
Furthermore, \cite{phillips1988asymptotic} proved the following limit result
\begin{align}
\int_0^1 W_{\epsilon} (r) dW_v (r) \Rightarrow \mathcal{N} \left( 0, \int_0^1 W_{\epsilon} (r) dr   \right)
\end{align}
Therefore, under the null hypothesis $\mathbb{H}_0: \beta = 0$, it follows that
\begin{align}
T \hat{\beta} \Rightarrow \left\{ \phi \frac{ \sigma^2_{ \epsilon } }{2}   \bigg[  W^2_{\epsilon} (1) + 1 \bigg]  +   \sigma_{ \epsilon } \sigma_{ v } \bigg[ \int_0^1 W_{\epsilon} (r) dW_v (r) \bigg] \right\} \left( \sigma^2_{ \epsilon } \int_0^1 W_{ \epsilon } (r)^2 dr \right)^{-1}.
\end{align}

%%-------------------------------------------------------------------------%%
\newpage 

Thus, the $t-$statistic, denoted as $\mathcal{T}_{\beta = 0}$ for testing the null hypothesis, $\mathbb{H}_0: \beta = 0$, is written as 
\begin{align}
\mathcal{T}_{\beta = 0} 
= \frac{ \widehat{\beta} }{  \widehat{\sigma}^2_u \left( \displaystyle \sum_{t=1}^T x_t^2 \right)^{- \frac{1}{2} }} 
= T \frac{ \widehat{\beta} }{  \widehat{\sigma}^2_u \left( \displaystyle T^{-2} \sum_{t=1}^T x_t^2 \right)^{- \frac{1}{2} }} 
\end{align}
has the following asymptotic distribution
\begin{align}
\mathcal{T}_{\beta = 0}  
&\Rightarrow
\left\{ \phi \frac{ \sigma^2_{ \epsilon } }{2}   \bigg[  W^2_{\epsilon} (1) + 1 \bigg]  +   \sigma_{ \epsilon } \sigma_{ v } \bigg[ \int_0^1 W_{\epsilon} (r) dW_v (r) \bigg] \right\} \left( \sigma^2_{ \epsilon } \int_0^1 W_{ \epsilon } (r)^2 dr \right)^{-\frac{1}{2}} \times \frac{1}{ \sigma_u^2   }
\\
&\equiv 
\frac{ \phi }{ 2 } \frac{  \sigma_{ \epsilon } }{ \sigma_u  } \bigg[  W^2_{\epsilon} (1) + 1 \bigg] \left(  \int_0^1 W_{ \epsilon } (r)^2 dr \right)^{-\frac{1}{2}} + \frac{  \sigma_{ v } }{ \sigma_u  } \mathcal{N}(0,1).
\end{align}

\medskip

\begin{remark}
Notice that the derived limiting distribution of the Student-t statistic indicates that the $t-$ratio of $\widehat{\beta}$ does not follow a standard normal distribution unless $\phi = 0$; in which case the structure of the model implies that the regressor $x_t$ is exogenous for the estimation of the model parameter $\beta$ which is the main parameter of interest for inference purposes. In particular, when  $\phi \neq 0$ then the first term of the above limiting distribution gives rise to second-order or endogeneity bias, which although asymptotically negligible in estimating $\beta$ due to super consistency, can appear in finite-samples causing size distortions when obtaining the empirical size of the test.   
\end{remark}

\subsubsection{Predictive Regression Model}

Predictive regression models are extensively used in time series econometrics and the empirical finance literature for examining the stock return predictability puzzle as proposed by \cite{campbell2006efficient}. A standard predictive regression has the following econometric specification (see, \cite{kostakis2015robust})
\begin{align}
y_t &= \mu + \beta x_{t-1} + \epsilon_t, \ \ t = 1,...,T
\\
x_t &= \rho x_{t-1} + u_t 
\end{align}
The innovation sequence $(\epsilon_t,u_t)$ is generated such that $(\epsilon_t,u_t)    \sim_{ \textit{i.i.d} } \mathcal{N} (0,\Sigma)$ where $\Sigma = 
\begin{bmatrix}
 \sigma^2_{\epsilon} & \sigma_{\epsilon u}  \\
 \sigma_{\epsilon u} & \sigma^2_{u}
\end{bmatrix}$. 

Similarly to the previous example, the null hypothesis of interest using linear restrictions on the model parameter $\beta$, is formulated such that $\mathbb{H}_0: \beta = 0$. The main econometric challenges when conducting statistical inference using the predictive regression model includes the problem of embedded endogeneity due to the innovation structure of the system as well as the nuisance parameter of persistence, $c$, when the autocorrelation coefficient of the model is expressed with  the local-to-unity specification. As a result, depending on the value of the autocorrelation coefficient, the asymptotics for the parameter of the predictive regression model take a different form which makes statistical inference challenging. Specifically, when $|\rho| < 1$, then $x_t$ is known to be stationary, when $\rho = 1$ then $x_t$ is unit root or integrated and when $c<0$ is assumed to follow a local-to-unity or nearly integrated process. The literature has proposed various methodologies for conducting statistical inference robust to the nuisance parameter of persistence. For instance, \cite{Phillips2007limit} study the limit theory of time series models which includes regressors that are close to the unit root boundary\footnote{The authors consider the limit distribution theory in both the near-stationary $(c < 0)$ and the near-explosive cases $(c > 0)$.}.

%%-------------------------------------------------------------------------%%
\newpage 

In terms of the asymptotic theory that correspond to the predictive regression model, we consider the partial-sum process for the integrated regressor. The particular aspect is important especially in comparison to when constructing residual-based statistics in which case the main quantity of interest is the partial-sum process of the residuals corresponding to stationary innovations. Denote with $\mathcal{F}_{T,t-1}$ the $\sigma-$algebra generated by the random variables and with $X_{ \floor{Tr} }$ the partial-sum process of interest. Then, under the assumption that the $x_t$ is generated as a local-unit-root process the weakly convergence of the partial-sum functional corresponds to a uniform convergence to an \textit{Ornstein-Uhlenbeck} (OU) process\footnote{The continuous time OU diffusion process given by $dy_t = \theta y_t dt + \sigma dw_t, \ y_0 = b, t > 0
$, where $\theta$ and $\sigma >0$ are unknown parameters and $w_t$ is the standard Wiener process, has a unique solution to $\{ y_t \}$ which is expressed as $ y_t = \text{exp} \left( \theta t  \right) b + \sigma \int_0^t \text{exp} \left[ \theta( t - s)  \right] dw_s \equiv \text{exp} \left( \theta t \right) + \sigma J_{\theta} \left( t \right)$ (see, e.g., \citep{perron1991continuous}).} rather to the standard Wiener process (see, \cite{Phillips1987time} and related limit theory in \cite{durrett1978functional}) as   
\begin{align}
\frac{1}{ \sqrt{T} } X_{ \floor{ Tr} } \Rightarrow J_c(r) := \int_0^r e^{(r-s)c} d B_c(s), \ r \in [0,1].       
\end{align}
In other words, the assumptions we impose regarding the parametrization of the autocorrelation coefficient $\rho$ can change the asymptotic behaviour of the stochastic difference equation. Generally, statistical inference is nonstandard in the sense that when $\rho_T = \left( 1 + \frac{c}{T} \right)$ for some nuisance parameter $c$, then the testing problem concerning the parameter $\beta$ exhibit nonstandard large-sample properties under local-to-unity asymptotics.  

\subsubsection{Predictive tests}

To provide some further clarity regarding the effect of expressing the autocorrelation coefficient in terms of moderate deviations from unity, to the validity of  conventional inference methods, we consider as an example the stationary autoregressive model AR(1), $y_t = \rho_T y_{t-1} + u_t$ where $u_t \overset{ i.i.d }{ \sim }(0, \sigma^2)$ and $| \rho | < 1$. In this case, it is a well-known fact that the limit distribution of the t-test for testing the null hypothesis $\mathbb{H}_0: \rho_T = 0$ with $\mathcal{T}_T ( \rho_T)  = \displaystyle \frac{ \widehat{\rho}_T - \rho }{ \widehat{\sigma} } \Rightarrow \mathcal{N}(0,1)$ converges to a standard normal distribution. In addition, the asymptotic distribution of $\mathcal{T}_T ( \rho_T)$ is invariant even under the assumption of conditional heteroscedasticity which implies $\mathbb{E}\left( u_t^2 | \mathcal{F}_{t-1}  \right) = \sigma^2_t$ and $\underset{ t \in \mathbb{Z} }{ \text{sup} } | \hat{\sigma}^2_t - \sigma^2_t | = o_p(1)$, a condition for consistent estimation. 

On the other hand, the limiting distribution of the model parameter $\beta$ of the predictive regression model as well as the associated t-test for testing the null hypothesis, $\mathbb{H}_0: \beta = 0$, appears to be challenging due to the fact that it is found to be nonstandard and the corresponding t-test is non-pivotal since it depends on the nuisance parameter $c$ (see, \cite{cavanagh1995inference} and \cite{campbell2006efficient}). Consequently, given the focus of our study to the asymptotic behaviour of  partial-sum processes when constructing test statistics in nonstationary time series models, we illustrate the related asymptotic theory with some examples.    

It is worth mentioning that the partial-sum processes, $X_{ \floor{Tr}} (r)$, are considered to be (maximally) invariant with respect to the presence of the model intercept $\mu$. Therefore under the null hypothesis, $\mathbb{H}_0: \beta = 0$, joint weak convergence of observation processes to their Brownian motion counterparts holds. Specifically, an application of the invariance principle proposed by \cite{Phillips1987time} such that $\frac{x_{ \floor{Tr} }}{\sqrt{T} } \Rightarrow J_c(r)$, where $J_c(r) = \int_{0}^r e^{ (r-s)c} dB_c(s)$ is a standard Ornstein-Uhlenbeck process, implies that
\begin{align}
\frac{1}{ T \sqrt{T} } \sum_{t=1}^{ \floor{Tr} } x_t    & \Rightarrow \int_0^r J_c(s) ds, 
\\
\frac{1}{T} \sum_{t=1}^{\floor{Tr} } x_t \epsilon_t  & \Rightarrow \int_0^r J_c(s) dB_{\epsilon}(s). 
\end{align}

%%-------------------------------------------------------------------------%%
\newpage 

Since $T \left( \widehat{\beta}_T - \beta  \right) = \mathcal{O}_p(1)$ is bounded in probability, then we can establish the usual mode of converges in distribution in the same probability space such that 
\begin{align}
T \left( \widehat{\beta}_T - \beta \right) 
= 
\frac{ \displaystyle \frac{1}{T} \sum_{t=1}^T x_{t-1} \epsilon_t  }{ \displaystyle \frac{1}{T^2} \sum_{t=1}^T x_{t-1}^2  }  \Rightarrow 
\frac{ \displaystyle \int_{0}^1 J_c(s) dB_{\epsilon}(s) }{ \displaystyle \int_{0}^1 J^2_x(s) ds  }.
\end{align}
Consider the stationary case such that $| \rho | < 1$, then by partitioning the covariance matrix $\Sigma$ similar to the regression model with the integrated regressor, we use the decomposition $\epsilon_{1.2 t} = \epsilon_t  - \frac{ \sigma_{\epsilon u} }{ \sigma_{\epsilon} } u_t$  which implies  
\begin{align} 
\label{termmixed}
T \left( \widehat{\beta}_T - \beta  \right) 
= 
\frac{ \displaystyle \frac{1}{T} \sum_{t=1}^T x_{t-1} \epsilon_t  }{ \displaystyle \frac{1}{T^2} \sum_{t=1}^T x_{t-1}^2  } 
= 
\frac{ \displaystyle \frac{1}{T} \sum_{t=1}^T x_{t-1} \epsilon_{1.2 t}  }{ \displaystyle \frac{1}{T^2} \sum_{t=1}^T x_{t-1}^2  } 
+ 
\frac{ \sigma_{\epsilon u} }{ \sigma_{\epsilon} }  \frac{ \displaystyle \frac{1}{T} \sum_{t=1}^T x_{t-1} u_t  }{ \displaystyle \frac{1}{T^2} \sum_{t=1}^T x_{t-1}^2  } 
\end{align}
The first term of expression (\ref{termmixed}) since in includes a conditional error term then the corresponding limit distribution converges to a mixed normal limit. Moreover, we consider the joint convergence of the martingale sequences $\left\{ \sum_{t=1}^n x_{t-1} \epsilon_t \right\}$ and $\left\{ \sum_{t=1}^n  u_t \right\}$ are defined on the same probability space
\begin{align}
\xi_{Tt} = \left( \frac{1}{T} x_{t-1} \epsilon_t, \frac{1}{\sqrt{T}}  u_t   \right)^{\prime}
\end{align}
Thus, the conditional covariance matrix of the martingale vector $\xi_{Tt}$ is given by 
\begin{align}
\sum_{t=1}^T \mathbb{E} \bigg( \xi_{Tt} \xi_{Tt}^{\prime} | \mathcal{F}_{t-1} \bigg) 
=
\begin{bmatrix}
\displaystyle \left( \frac{1}{T^2} \sum_{t=1}^T x_{t-1}^2 \right) \sigma^2_{\epsilon} &  \displaystyle \left( \frac{1}{ T \sqrt{T} } \sum_{t=1}^T x_{t-1} \right) \sigma_{ \epsilon u} \\
\displaystyle \left( \frac{1}{ T \sqrt{T} } \sum_{t=1}^T x_{t-1} \right) \sigma_{ u \epsilon }   & \displaystyle \sigma_{uu}
\end{bmatrix}
\end{align}
Using the partition matrix identity, $\Sigma_{1.2} = \Sigma_{11} - \Sigma_{12} \Sigma_{22}^{-1} \Sigma_{21}$, to the predictive regression model we obtain the relation $\frac{ \displaystyle \sigma_{1.2} }{ \displaystyle \sigma^2_{\epsilon} } =  1 - \frac{ \displaystyle \sigma^2_{ u \epsilon } }{ \displaystyle \sigma^2_{\epsilon} \sigma^2_{u} }$. Therefore, the following mixed normal limit convergence holds
\begin{align}\label{mixed}
\frac{ \displaystyle \frac{1}{T} \sum_{t=1}^T x_{t-1} \epsilon_{1.2 t}  }{ \displaystyle \frac{1}{T^2} \sum_{t=1}^T x_{t-1}^2  } 
\Rightarrow 
\mathcal{MN} \left( 0, \sigma_{1.2} \left(  \int_{0}^1  J_c(r) dr \right)^{-1} \right)
\end{align}  
Hence, using expressions (\ref{termmixed}) and (\ref{mixed}) and the limit result below 
\begin{align}
T \widehat{ \sigma}_{\beta} = \widehat{\sigma}_{\epsilon} \left( \frac{1}{T^2} \sum_{t=1}^T x^2_{t-1} \right)^{-1/2} \Rightarrow  \sigma_{\epsilon} \left( \int_0^1 J^2_c(r) dr  \right)^{-1}
\end{align}
for some $r \in (0,1)$, we obtain an analytical expression for the asymptotic distribution of the t-statistic $\mathcal{T}_T ( \beta_T ) = \frac{ \widehat{\beta}_T - \beta }{ \widehat{ \sigma}_{\beta} }  \Rightarrow \phi \widehat{\mathcal{M}}(c) + (1 - \phi^2)^{1/2} \mathcal{Z}$, where $\mathcal{Z} \sim \mathcal{N}(0,1)$ is independent of the random quantity $\widehat{\mathcal{M}}(c)$ and $c$ denotes the nuisance parameter of persistence.

%%-------------------------------------------------------------------------%%
\newpage 

More precisely, it holds that 
\begin{align*}
\mathcal{T}_T ( \beta_T ) 
= 
\frac{ \widehat{\beta}_T - \beta }{ \widehat{ \sigma}_{\beta} } 
\Rightarrow 
\frac{1}{ \displaystyle \sigma_{\epsilon} \left(  \int_0^1 J^2_x(r) dr  \right)^{-1} } 
\left\{ \displaystyle \mathcal{MN} \left(  0,   \sigma_{1.2} \left(  \int_{0}^1  J_c(r) dr \right)^{-1} \right) +  \frac{ \sigma_{\epsilon u} }{ \sigma_{\epsilon} }  \frac{ \displaystyle \int_{0}^1 J_c(s) dB_{x}(s) }{ \displaystyle \int_{0}^1 J^2_c(s) ds}  \right\},
\end{align*}
where $\phi =  \frac{ \displaystyle \sigma_{ \epsilon u } }{ \displaystyle \sigma_{\epsilon} \sigma_{u}}$, and $\widehat{\mathcal{M}}(c) =  \frac{  \displaystyle \int_{0}^1 J_c(s) dB_{x}(s) }{ \displaystyle \sigma^2_{u}  \int_{0}^1 J^2_c(s) ds}$.

In summary, the t-statistic for the predictive regression coefficient has been proved to have a non-standard limiting distribution which implies that normal or chi-square based inference is not available in practise, due to the endogeneity problem as well as the existence of persistence regressors. Therefore, the particular non-standard testing problem makes it difficult to conduct inference without prior knowledge regarding the exact value of the coefficient of persistence and in practise cannot be consistently estimated. Suggested solutions to overcome this problem include the Bonferroni confidence interval proposed by \cite{cavanagh1995inference} and \cite{elliott1996efficient}, the conditional likelihood approach that uses sufficient statistics proposed by \cite{jansson2006optimal} and the control function approach proposed by \cite{elliott2011control}.

\subsection{Structural Break Testing}

Setting against the background described in details in Section 4.1 we now discuss the main challenges for the development of the structural break testing framework. 

\subsubsection{Asymptotic Distributions of Test Statistics}

In this section we consider the implementation of OLS-CUSUM and Wald-type statistics within the local-to-unity framework when detecting instabilities in the parameters of a predictive regression model with persistent regressors. Although in this paper we consider an in-sample monitoring scheme, under suitable modifications an on-line (sequential) monitoring scheme as proposed by \cite{chu1996monitoring} can provide an early warning mechanism for risk management purposes based on macroeconomic and financial conditions. The implementation of such a framework can be interpreted as a dynamic methodology for testing for parameter instability under the assumption of time-varying persistence properties. 

Therefore, we are interested in proposing suitable testing methodologies for detecting structural change in the vector of regression coefficients $\theta$ of the following time series regression model 
\begin{align}
y_t = x_{t-1}^{\prime} \beta_1 \mathbf{1} \{ t \leq k \}  + x_{t-1}^{\prime} \beta_2 \mathbf{1} \{ t > k \} + \epsilon_t, \ \ \ t = 1,...,T  ,
\end{align}
where $x_t = R_T x_{t-1} + u_t$ with $R_T = \left( 1 - \frac{C}{T} \right)$ and $C = \mathsf{diag} \left\{ c_1,..., c_p \right\}$ (see, \cite{kostakis2015robust}). Under the null hypothesis of no structural break $\mathbb{H}_0: \beta_1 = \beta_2$. Our proposition aims to incorporate neglected non-linearities such as structural breaks to the current LUR framework. For example, certain non-linear functions\footnote{For example, \cite{wang2012specification} consider a specification test for nonlinear nonstationary models within the LUR framework of cointegrating regression system.} of $I(1)$ processes can wrongly behave like stationary long memory processes (see, e.g., \cite{kasparis2014nonlinearity}). A recent approach which considers structural breaks under such conditions is presented by \cite{berenguer2020cumulated}. In this paper, we consider the weak dependence assumption. 

%%-------------------------------------------------------------------------%%
\newpage 

\subsubsection{OLS-CUSUM test statistic}

Next, we focus on the limit theory of the residual-based statistic, CUSUM test, constructed using the OLS residuals of the predictive regression under the null hypothesis, $\mathbb{H}_0: \beta_1 = \beta_2$. The OLS residuals of the predictive regression are given by  
$\widehat{ \epsilon }_t^{ ols } = y_t - x_t^{\prime} \widehat{\beta}^{ ols }$ and the corresponding OLS-CUSUM statistic is expressed in the usual way as below for $s \in [0,1]$
\begin{align}\label{ivxcusum}
\mathcal{\widetilde{C}}_T(k) 
= 
\underset{ 0 \leq k \leq T }{ \mathsf{max} }  \  
\frac{ \displaystyle \sum_{t=1}^{k} \widehat{\epsilon}_t^{ ols } - \frac{k}{T} \sum_{t=1}^T \widehat{\epsilon}_t^{ ols} }{ \displaystyle \widehat{\sigma}_{\epsilon} \sqrt{T} }  
=  
\underset{ 0 \leq r \leq 1 }{ \mathsf{sup} } \ \frac{ \displaystyle \frac{1}{\sqrt{T}} \sum_{t=1}^{ \floor{Tr} } \widehat{\epsilon}_t^{ ols } - \frac{r}{\sqrt{T}} \sum_{t=1}^T \widehat{\epsilon}_t^{ols} }{ \widehat{\sigma}_{\epsilon} \sqrt{T} } 
\end{align}
The OLS residuals of the predictive regression can be expressed as $\widehat{ \epsilon }_t^{ ols } = y_t - x_t^{\prime} \widehat{\beta}^{ ols } \equiv \epsilon_t - x_t^{\prime} \left( \widehat{\beta}_T - \beta  \right) $ 
\begin{align}
\frac{1}{\sqrt{T} } \sum_{t=1}^{ \floor{Tr} }  \widehat{ \epsilon }_t 
&= 
\frac{1}{ \sqrt{T} } \sum_{t=1}^{ \floor{Tr} } \epsilon_t - T \left( \widehat{\beta}_T - \beta \right) \frac{1}{ T \sqrt{T} } \sum_{t=1}^{ \floor{Tr} } x_t  
\\
\frac{1}{\sqrt{T} } \sum_{t=1}^{ \floor{Tr} }  \widehat{ \epsilon }_t 
&\Rightarrow_d 
B_{\epsilon}(r) - \left\{ \frac{ \displaystyle \int_0^r J_c (s) dB_u(s) }{ \displaystyle \int_0^1 J_c^2(s) ds  } \right\} \times \int_{0}^r J_c (s) ds .
\end{align}
Therefore, using expression (\ref{ivxcusum}) the OLS-CUSUM statistic within the LUR framework becomes
\begin{align}
\mathcal{C}_T(k) 
&= 
\frac{1}{ \widehat{\sigma}_{\epsilon} } \left\{ \left(  B_u(r) - \frac{ \displaystyle \int_0^r J_c (s) dB_u(s) }{ \displaystyle \int_0^1 J_c^2(s) ds  } \int_{0}^r J_c (s) ds  \right) 
- 
r \left( B_u(1) - \frac{ \displaystyle \int_0^r J_c (s) dB_u(s) }{ \displaystyle \int_0^1 J_c^2(s) ds  } \int_0^1 J_c (s) ds \right)   \right\} 
\nonumber
\\
\nonumber
\\
\mathcal{C}_T(k)
&= 
\frac{1}{ \widehat{\sigma}_{\epsilon} } \left\{  \bigg[ B_u(r) - r B_u(1) \bigg] -  \frac{ \displaystyle  \int_0^r J_c (s) dB_u(s) }{ \displaystyle \int_0^1 J_c^2(s) ds  } \times \left( \int_0^r J_c (s) ds  -  r \int_0^1 J_c (s) ds   \right)  \right\}.
\end{align}

As we can clearly observe from the second term of the above expression that corresponds to the limiting distribution of the OLS-CUSUM statistic in a predictive regression model with persistent regressors the dependence on the nuisance parameter of persistence, $c$, makes the limit result non-standard and non-pivotal. In other words, the implementation of a residual-based statistic in a predictive regression model using OLS residuals is considered to be problematic in the derivation of the asymptotic distribution due to its dependence on the nuisance degree of persistence of the autoregressive specification of the model. Furthermore, for $k = \floor{Tr}$ we define the following term for notation simplicity 
\begin{align}
\widetilde{J}_{\infty }( c ; r) := \frac{ \displaystyle  \int_0^r J_c (s) dB_u(s) }{ \displaystyle  \int_0^1 J_c^2(s) ds  } \times \int_0^r J_c (s) ds. 
\end{align}
Then, the weakly convergence result can be written as below 
\begin{align}
\mathcal{C}_T(k) \Rightarrow \left( \bigg[ W(r)  - r W(1) \bigg] - \bigg[ \widetilde{J}( c ; r) - r \widetilde{J}( c, 1) \bigg]  \right).
\end{align}

%%-------------------------------------------------------------------------%%
\newpage 

In summary, the weakly convergence of the in-sample  OLS-CUSUM statistic includes the component $\widetilde{J}_{\infty} ( c, r)$ which depends on the nuisance parameter $c$ and thus can affect the true size of the test under the null hypothesis of no parameter instability given the fact that we cannot consistently estimate the coefficient of persistence. However, for example the term $\int_0^r J_c(s) ds - r \int_0^1 J_c(s) ds$ it is likely to be quite small and therefore can be considered not to be contributing to huge size distortions. An extensive Monte Carlo study can shed light on the particular aspect of the proposed test for detecting structural change in predictive regression models with persistent regressors. Thus, we have demonstrated that when testing for a structural break in linear time series regression models, the partial-sum processes of conventional test such as those of residual-based and Wald-type statistics have different properties when information regarding the integration order of regressors is available in the form of the LUR specification form.

\section{CONCLUSION}
\label{Section5}

In this paper we establish the Brownin Bridge limiting distributions in a fairly standard settings, that is linear regression models under the assumption of stationarity and ergodicity when constructing  residual-based and Wald-type statistics for testing the null hypothesis of no parameter instability. In particular, in all those cases we have demonstrated that the normalized Brownian bridge limit holds for both test statistics. Additionally, we investigate whether this property also holds for nonstationary time series regression models with integrated or persistent regressors. Our asymptotic theory analysis has demonstrated that while in the classical linear regression model with stationary regressors the convergence of the test statistics to  brownian bridge limit results hold, in the case of the nonstationary time series model it appears to be the case that the limiting distribution is non-standard and non-pivotal due to the dependence of the distribution to the nuisance parameter of persistence.

Based on the general assumption that the underline stochastic processes are mean-reverting, then we can establish adequate approximations to finite-sample moments for the model under consideration regardless of the econometric environment operates under the assumption of stationarity or we consider the settings of a nonstationary time series model. On the other hand, since the main feature of nonstationary time series models is the parametrization of the autocorrelation coefficient with respect to the nuisance parameter of persistence, this implies that limiting distributions are non-standard and non-pivotal which makes inference difficult. In particular, despite the large availability of macroeconomic and financial variables which can be included as regressors in predictive regressions (such as financial ratios, diffusion indices, fundamentals), practitioners have no prior knowledge regarding the persistence properties of predictors so conventional estimation and inference methods for model parameters, such as predictability tests, forecast evaluation tests as well as structural break testing require to handle the nuisance parameter of persistent.

Therefore, further research is needed to propose suitable statistical methodologies that take into consideration these challenges, especially when testing for the presence of a structural break in nonstationary time series models. Practically, the extension of structural break tests in nonstationary time series models, such as predictive regressions, which are particularly useful when information regarding the time series properties of regressors is not lost by taking the first difference for instance, is crucial for both theoretical and empirical studies. In a subsequent paper, we propose a formal econometric framework and develop the associated asymptotic theory for Wald-type statistics under regressors nonstationarity.

%%-------------------------------------------------------------------------%%
\newpage

\section{APPENDIX}

\subsection{Appendix A: Technical Proofs}

\paragraph{Proof of Theorem  \ref{theorem2}}

\begin{proof}
Consider the functional $\widehat{\sigma}^2_{z}  \mathcal{Z}_T(s)$ given by
\begin{align*}
\widehat{\sigma}^2_{z} \mathcal{Z}_T(s) = T \left( \bar{y_{1}} - \bar{y_{2}} \right)^2 
= 
T \left( \frac{1}{ \floor{Ts} } \sum_{ t = 1 }^{ \floor{Ts} } y_t -  \frac{1}{ T - \floor{Ts} } \sum_{ t = \floor{Ts} + 1 }^{ T } y_t  \right)^2. 
\end{align*}
Under the null hypothesis $\mathbb{H}_0: \theta_1 = \theta_2$, we have that the following expression holds
\begin{align*}
\left( \bar{y_{1}} - \bar{y_{2}} \right) &= \left( \frac{1}{k} \sum_{t=1}^k \epsilon_t - \frac{1}{T-k} \sum_{t=k+1}^T \epsilon_t  \right) 
= \frac{1}{k} \sum_{t=1}^k \epsilon_t  - \frac{1}{T-k} \left( \sum_{t=1}^T \epsilon_t - \sum_{t=1}^k \epsilon_t \right) \\
&= \frac{T \sqrt{T}}{ k( T-k) }   \left( \sum_{t=1}^k \frac{ \epsilon_t }{ \sqrt{T} } -  \frac{k}{T} \sum_{t=1}^T \frac{ \epsilon_t }{ \sqrt{T} } \right)
\end{align*}
Therefore we obtain that 
\begin{align}
\label{proof}
\widehat{\sigma}^2_{z} \mathcal{Z}_T(s) 
=
T \left( \bar{y_{1}} - \bar{y_{2}} \right)^2 
= 
\frac{1}{ \left( \frac{k}{T} \right)^2 \left( 1 - \frac{k}{T} \right)^2 } \left( \sum_{t=1}^k \frac{ \epsilon_t }{ \sqrt{T} } -  \frac{k}{T} \sum_{t=1}^T \frac{ \epsilon_t }{ \sqrt{T} } \right)^2. 
\end{align}   
Thus, equality (\ref{proof}) shows that $\widehat{\sigma}^2_{z} \mathcal{Z}_T(s) \overset{\text{plim}}{=} \frac{1}{s^2(1-s)^2} \widehat{\sigma}^2_{\epsilon} \left[  \mathcal{Z}^o_T (s) \right]^2$, due to the continuous mapping theorem. Since $\widehat{\sigma}^2_{\epsilon} =  \widehat{\sigma}^2_{z} \times s (1-s)$ and using the weakly convergence result for $\mathcal{Z}^o_T(s)$ deduced in Example \ref{Example2} (see, \cite{aue2013structural}), which implies that $ \mathcal{Z}^o_T(s) \Rightarrow \underset{ s \in [\nu , 1 - \nu]  }{ \text{sup} } \mathcal{BB}(s)$ gives the following asymptotic result for the test statistic $\mathcal{Z}_T (s)$ provided that $0 < \nu < 1$
\begin{align}
\mathcal{Z}_T (s)  \Rightarrow \underset{ s \in [\nu, 1 - \nu]  }{ \text{sup} } \frac{  \big[ W(s) - s W(1) \big]^2 }{ s(1 - s) }.
\end{align} 
\end{proof}

\paragraph{Proof of Theorem  \ref{theorem3}}

\begin{proof}
Let $\widehat{\theta}_{1}$ and $\widehat{\theta}_{2}$ be the OLS estimator for  $\{1,..., k \}$ and $\{ k + 1,...,T \}$ respectively, then it follows that 
\begin{align*}
\widehat{\theta}_{1} 
&= 
\left(  \frac{1}{T} \sum_{t=1}^{k} x_{1,t} x_{1,t}^{\prime}  \right)^{-1}    \left(  \frac{1}{T} \sum_{t=1}^{k} x_{1,t} y_t  \right)   
\\
\widehat{\theta}_{2} 
&=  
\left(  \frac{1}{T} \sum_{t=k + 1}^{T} x_{2,t} x_{2,t} \right)^{-1} \left( \frac{1}{T} \sum_{t = k + 1}^{T} x_{2,t}^{\prime} y_t  \right) 
\end{align*}

%-------------------------------------------------------------------------%%
\newpage

Using the restriction matrix and the coefficient matrix we formulate the Wald statistic as below 
\begin{align*}
\mathcal{W}_T(s) &= \frac{1}{ \widehat{\sigma}^2_{\epsilon} } \left( \widehat{\theta}_1 - \widehat{\theta}_1 \right)^{\prime} \left[ \left( X_1^{\prime} X_1 \right)^{-1} +  \left( X_2^{\prime} X_2 \right)^{-1} \right]^{-1} \left( \widehat{\theta}_1 - \widehat{\theta}_1 \right)
\end{align*}

Let $A$ and $B$ be matrices whose dimensions are such that the addition $\left ( A + B \right)$ is well-defined, we can use the binomial matrix identity
\begin{align*}
\left( A + B \right)^{-1} = A^{-1} - A^{-1}\left( B^{-1} + A^{-1} \right)^{-1} A^{-1}
\end{align*}
and since the following property holds due to the orthogonality condition (and no model intercept)
\begin{align*}
\left(   X^{\prime} X  \right) = \left( X_1^{\prime} X_1 \right) +  \left( X_2^{\prime} X_2 \right)
\end{align*}
then we get that 
\begin{align*}
\left[ \left( X_1^{\prime} X_1 \right)^{-1} +  \left( X_2^{\prime} X_2 \right)^{-1} \right] = \left[ X_1^{\prime}X_1 - X_1^{\prime}X_1  \left( X^{\prime}X \right)^{-1} X_1^{\prime}X_1   \right]
\end{align*}
Therefore, keeping the matrix notation for simplicity the Wald test is expressed as below
\begin{align*}
\mathcal{W}_T(s) = \frac{1}{ \widehat{\sigma}^2_{\epsilon} } \left( \widehat{\theta}_1 - \widehat{\theta}_1 \right)^{\prime} \left[ X_1^{\prime}X_1 - X_1^{\prime}X_1  \left( X^{\prime}X \right)^{-1} X_1^{\prime}X_1   \right]^{-1} \left( \widehat{\theta}_1 - \widehat{\theta}_1 \right)
\end{align*}
which according to \cite{pitarakis2008comment} has an equivalent representation given by 
\begin{equation*}
\mathcal{W}_T(s) = \frac{1}{ \hat{\sigma}^2_{\epsilon} } \Big[ \epsilon^{\prime} X_1 - \epsilon^{\prime} X(X^{\prime}X)^{-1} X_1^{\prime}X_1 \Big]  \left[ X_1^{\prime}X_1 - X_1^{\prime}X_1  \left( X^{\prime}X \right)^{-1} X_1^{\prime}X_1   \right] ^{-1} \Big[ X_1^{\prime} \epsilon -  X_1^{\prime}X_1 \left( X^{\prime}X \right)^{-1} X^{\prime} \epsilon  \Big]
\end{equation*}
Then, the econometric specification of the model with when testing for a single structural break at an unknown break- point $k = \floor{Ts}, s \in (0,1)$ is given by 
\begin{align}
y = X_1 \theta_1 + X_2 \theta_2 + \epsilon \overset{\mathbb{H}_0}{ = } X \theta_1 + \epsilon
\end{align}
Then, the OLS estimators are expressed in matrix form as below
\begin{align*}
\widehat{\theta}_1 &= \left( X_1^{\prime} X_1 \right)^{-1}  X_1^{\prime} y \ \ \ \text{and} \ \ \ \widehat{\theta}_2 = \left( X_2^{\prime} X_2 \right)^{-1}  X_2^{\prime} y  
\end{align*}
Under $\mathbb{H}_0: \theta_1 = \theta_2$, we have that $y = X_1 \theta_1 + X_2 \theta_1 + \epsilon = X \theta_1$, where $X = X_1 + X_2$ since $X_1 \perp X_2$. Thus, $X_1^{\prime} X = X_1^{\prime} X_1$ and $X_1^{\prime} X_2 = X_2^{\prime} X_1 = 0$ due to the orthogonality property. Therefore, the OLS estimators are written in the following form 
\begin{align*}
\widehat{\theta}_1 &= \left( X_1^{\prime} X_1 \right)^{-1} X_1^{\prime} \left( X \theta_1 + \epsilon \right) =  \theta_1 + \left( X_1^{\prime} X_1 \right)^{-1} \left( X_1^{\prime}  X_2 \right)\theta_1 +  \left( X_1^{\prime} X_1 \right)^{-1} X_1^{\prime} \epsilon   
\\
\widehat{\theta}_2 &= \left( X_2^{\prime} X_2 \right)^{-1} X_2^{\prime} \left( X \theta_1 + \epsilon \right) = \theta_1 + \left( X_2^{\prime} X_2 \right)^{-1} \left( X_2^{\prime}  X_1 \right)\theta_1 +  \left( X_2^{\prime} X_2 \right)^{-1} X_2^{\prime} \epsilon
\end{align*}
Consider the expression
\begin{align*}
\sqrt{T} \left( \widehat{\theta}_1 - \widehat{\theta}_2 \right) 
&=  \left[ \left( X_1^{\prime} X_1 \right)^{-1} X_1^{\prime} \epsilon  - \left( X_2^{\prime} X_2 \right)^{-1} X_2^{\prime} \epsilon \right] \\
&\equiv  
\left\{ \left( \frac{1}{T} \sum_{t=1}^k x_{1,t} x_{1,t}^{\prime} \right)^{-1} \frac{1}{\sqrt{T}} \sum_{t=1}^k x_{1,t} \epsilon_t  -  \left( \frac{1}{T} \sum_{t=k+1}^T x_{2,t} x_{2,t}^{\prime} \right)^{-1}   \frac{1}{\sqrt{T}} \sum_{t=k+1}^T  x_{2,t} \epsilon_t \right\}
\end{align*}

%\Big[ \left( X_1^{\prime} X_1 \right)^{-1} X_1^{\prime} \epsilon - \left( X^{\prime}X  - X_1^{\prime} X_1  \right)^{-1} \left( X^{\prime} - X_1^{\prime} \right)\epsilon    \Big] \\
%&= \Big[  \left\{ \left( X_1^{\prime} X_1 \right)^{-1} + \left( X^{\prime}X  - X_1^{\prime} X_1  \right)^{-1}   \right\} X_1^{\prime} \epsilon   - \left( X^{\prime}X  - X_1^{\prime} X_1  \right)^{-1}  X^{\prime} \epsilon  \Big]

%%-------------------------------------------------------------------------%%
\newpage 

Furthermore, the following probability limits hold uniformly for $s \in (0,1)$,
\begin{align}
\left( \frac{1}{T} \sum_{t=1}^{ \floor{Ts} }     x_{1,t} x_{1,t}^{\prime} \right)  \overset{ p}{\to} s \boldsymbol{Q}_T, \ \ \left( \frac{1}{T} \sum_{t=\floor{Ts} + 1}^{T} x_{2,t} x_{2,t}^{\prime} \right) \overset{ p}{\to} (1-s) \boldsymbol{Q}_T, \ \ \left( \frac{  X^{\prime} X }{T} \right) \overset{ p}{\to} \boldsymbol{Q}_T 
\end{align}
which holds due to moment homogeneity. Furthermore, using the following two limit results
\begin{itemize}

\item[\underline{Result A:}]
\begin{align}
\left[ \left( \frac{ X_1^{\prime} X_1 }{T} \right)^{-1} +  \left( \frac{ X_2^{\prime} X_2 }{ T} \right)^{-1} \right]^{-1} \overset{ \text{plim}}{=} \left[  \frac{1}{s} \boldsymbol{Q}_T^{-1} + \frac{1}{1-s} \boldsymbol{Q}_T^{-1} \right]^{-1} = \left[ \frac{ \boldsymbol{Q}_T^{-1} }{ s(1-s) } \right]^{-1}
\end{align}

\item[\underline{Result B:}]

\begin{align}
\frac{1}{ \widehat{\sigma}^2_{\epsilon} }  \left( \frac{ X_1^{\prime}\epsilon }{\sqrt{T} }  - s \frac{ X^{\prime} \epsilon}{\sqrt{T} } \right)^{2} 
= 
\left( \frac{1}{ \widehat{\sigma}_{\epsilon} \sqrt{T}} \sum_{t=1}^{ \floor{Ts} } x_{t} \epsilon_t - s \frac{1}{ \widehat{\sigma}_{\epsilon} \sqrt{T}} \sum_{t=1}^{T} x_{t} \epsilon_t \right)^2 \Rightarrow  \big[ \boldsymbol{W}_p(s) - s \boldsymbol{W}_p(1) \big]^2
\end{align}
using the fact that $X_2^{\prime} \epsilon = \left( X^{\prime} \epsilon - X_1^{\prime} \epsilon  \right)$.
\end{itemize}

Therefore, we obtain the following limit result 
\begin{align*}
\sqrt{T} \left( \widehat{\theta}_1 - \widehat{\theta}_2 \right) 
& \overset{ \text{plim}}{=} 
\left\{  \frac{1}{s} \boldsymbol{Q}_T^{-1} \sigma_{\epsilon} \boldsymbol{W}_p(s)   -  \frac{1}{s(1-s)} \boldsymbol{Q}_T^{-1} \sigma_{\epsilon}  \big[ \boldsymbol{W}_p(1) - \boldsymbol{W}_p(s) \big] \right\} 
\\
&= 
\sigma_{\epsilon} \boldsymbol{Q}_T^{-1} \left\{ \frac{1}{s} \boldsymbol{W}_p (s) - \frac{1}{s(1-s)}   \big[ \boldsymbol{W}_p(1) - \boldsymbol{W}_p(s) \big] \right\} 
\\
&= 
\sigma_{\epsilon} \boldsymbol{Q}_T^{-1} \frac{1}{s(1-s)} \bigg\{ \boldsymbol{W}_p(s) - s \boldsymbol{W}_p(1)  \bigg\}
\end{align*}

Then, the limiting distribution of the Wald statistic can be expressed as below 
\begin{align*}
\mathcal{W}_T(s) 
&\Rightarrow 
 \frac{1}{ \sigma^2_{\epsilon} } \frac{\sigma_{\epsilon}}{s(1-s)} \bigg\{ \boldsymbol{W}_p(s) - s \boldsymbol{W}_p(1)  \bigg\}^{\prime} \boldsymbol{Q}_T^{-1} \left[ \frac{ \boldsymbol{Q}_T^{-1} }{ s(1-s) } \right]^{-1}  \frac{\sigma_{\epsilon}}{s(1-s)} \boldsymbol{Q}_T^{-1} \bigg\{ \boldsymbol{W}_p(s) - s \boldsymbol{W}_p(1)  \bigg\}
\\
\\
\mathcal{W}_T(s) 
&\Rightarrow  
\frac{  \bigg[ \boldsymbol{W}_p(s) - s \boldsymbol{W}_p(1) \bigg]^{\prime} \boldsymbol{Q}_T^{-1}   \bigg[ \boldsymbol{W}_p(s) - s \boldsymbol{W}_p(1) \bigg] }{s(1-s)} 
\end{align*} 

Therefore, the limiting distribution of the sup-Wald test follows
\begin{align}
\mathcal{W}^{*}_T(s) := \underset{ s \in [ \nu, 1 - \nu] }{ \mathsf{sup} } \mathcal{W}_T(s) \Rightarrow \underset{ s \in [ \nu, 1 - \nu] }{ \mathsf{sup} }  \frac{  \bigg[ \boldsymbol{W}_p(s) - s \boldsymbol{W}_p(1) \bigg]^{\prime} \boldsymbol{Q}_T^{-1}   \bigg[ \boldsymbol{W}_p(s) - s \boldsymbol{W}_p(1) \bigg] }{s(1-s)} 
\end{align}
which verifies a weakly convergence to a normalized squared Brownian bridge process.
\end{proof}

%%-------------------------------------------------------------------------%%
\newpage 

\begin{wrap}

\subsection{Appendix B: Supplementary Results}

Consider the regression model given by 
\begin{align}
y_t = X_t^{\prime} \beta + u_t
\end{align}

Define with $\eta_t = \big\{ X_t u_t \big\}$ to be a moment sequence and define with $\boldsymbol{\Omega} = \boldsymbol{\Lambda} \boldsymbol{\Lambda}^{\prime}$ where $\boldsymbol{\Lambda}$ is a lower triangular matrix based on the Cholesky decomposition of $\boldsymbol{\Omega}$. Notice that $\boldsymbol{\Omega}$ is equal to $2 \pi$ times the spectral density matrix of $\eta_t$ evaluated at frequency zero. Furthermore, define with $S_t = \sum_{j=1}^t \eta_t$ to be the partial-sum process for $\eta_t = \big\{ X_t u_t \big\}$.   

\medskip

Suppose that the regression above is estimated by OLS to obtain $\widehat{\beta}$, the OLS estimate. Then, the limiting distribution of $\sqrt{T} \left( \widehat{\beta}_T - \beta \right)$ is expressed as below
\begin{align*}
\sqrt{T} \left( \widehat{\beta}_T - \beta \right) 
&= 
\left( \frac{1}{T} \sum_{t=1}^T X_t X_t^{\prime} \right)^{-1} \left( \frac{1}{ \sqrt{T} } \sum_{t=1}^T X_t u_t \right)
\\
&=
\left( \frac{1}{T} \sum_{t=1}^T X_t X_t^{\prime} \right)^{-1} \frac{1}{ \sqrt{T} } S_T
\end{align*} 
Therefore, weak convergence result holds, $\sqrt{T} \left( \widehat{\beta}_T - \beta \right) \Rightarrow \boldsymbol{Q}^{-1} \boldsymbol{\Lambda} \boldsymbol{W}_k(1) \sim \mathcal{N} \left( \boldsymbol{0}, \boldsymbol{Q}^{-1} \boldsymbol{\Lambda} \boldsymbol{\Lambda}^{\prime} \boldsymbol{Q}^{-1} \right) \equiv \mathcal{N} \left( \boldsymbol{0}, \boldsymbol{V}  \right)$. In particular, the asymptotic normality property can be employed to test hypotheses about $\beta$. Thus, to construct standard tests that are asymptotically invariant to nuisance parameters, an estimate of $\boldsymbol{V} = \boldsymbol{Q}^{-1} \boldsymbol{\Omega} \boldsymbol{Q}^{-1}$ is required. Then, a natural estimator of $\boldsymbol{Q}^{-1}$ is given by $\left( T^{-1} \sum_{t=1}^T X_t X_t^{\prime} \right)^{-1}$. In particular, a HAC estimator of $\boldsymbol{\Omega}$ can be constructed from $\widehat{\eta} = X_t \widehat{u}_t$ where $\widehat{u}_t$ are the OLS residuals. 

Furthermore, consider the estimator $\widehat{\boldsymbol{V}}$ defined as below
\begin{align}
\widehat{\boldsymbol{V}} 
= 
\left( \frac{1}{T} \sum_{t=1}^T X_t X_t^{\prime} \right)^{-1} \widehat{\boldsymbol{\Omega}} \left( \frac{1}{T} \sum_{t=1}^T X_t X_t^{\prime} \right)^{-1}
\end{align} 
where $\widehat{\boldsymbol{\Omega}}$ is the HAC estimator of $\boldsymbol{\Omega}$. Moreover, if we can transform the expression $\sqrt{T} \left( \widehat{\beta}_T - \beta \right)$ using 
\begin{align}
\widehat{\boldsymbol{V}}^{- 1/ 2} = \left( \frac{1}{T} \sum_{t=1}^T X_t X_t^{\prime} \right)^{-1} \widehat{\boldsymbol{\Lambda}}
\end{align}
where $\widehat{\boldsymbol{\Lambda}}$ is obtained from the Cholesky decomposition of $\boldsymbol{\Omega}$, we have that 
\begin{align}
\widehat{\boldsymbol{V}}^{-1 /2} \sqrt{T} \left( \widehat{\beta}_T - \beta \right) \Rightarrow \mathcal{N} \big( \boldsymbol{0}_K, \boldsymbol{I}_K \big). 
\end{align} 

\medskip

\begin{remark}
Notice that the asymptotic theory does not explicitly account for the effect of sampling variation in $\widehat{\boldsymbol{V}}$, and this variation is potentially important in finite samples. However, our approach to deal with the particular aspect follows a similar transformation to the one we apply to the quantity $\sqrt{T} \left( \widehat{\beta}_T - \beta \right)$ using a data-driven moment matrix that does not require an estimate of the covariance matrix $\boldsymbol{\Omega}$.    
\end{remark}

%%-------------------------------------------------------------------------%%
\newpage 

Define with $\widehat{S}_t = \sum_{j=1}^t X_j \widehat{u}_j = \sum_{j=1}^t \widehat{\eta}$. Then, using the Assumptions, the limiting distribution of $T^{- 1 / 2} \widehat{S}_{ \floor{ T r} }$ as $T \to \infty$ is given by the following expression  
\begin{align*}
\frac{1}{ \sqrt{T} } \widehat{S}_{ \floor{ T r} } 
=
\frac{1}{ \sqrt{T} } \sum_{j=1}^t X_j \widehat{u}_j
&= 
\frac{1}{ \sqrt{T} } \sum_{j=1}^t \left[ X_t u_t - X_t X_t^{\prime} \left( \widehat{\beta}_T - \beta \right) \right]
\\
&=
\frac{1}{ \sqrt{T} } \sum_{j=1}^t \eta_t - \left( \frac{1}{T} \sum_{j=1}^t X_t X_t^{\prime} \right) \sqrt{T} \left( \widehat{\beta}_T - \beta \right) 
\\
&= 
\frac{1}{ \sqrt{T} } \sum_{j=1}^t S_{ \floor{ T r} } - \left( \frac{1}{T} \sum_{j=1}^t X_t X_t^{\prime} \right) \sqrt{T} \left( \widehat{\beta}_T - \beta \right) 
\\
&\Rightarrow \boldsymbol{\Lambda} \boldsymbol{W}_k(r) - r \boldsymbol{Q} \boldsymbol{Q}^{-1} \boldsymbol{\Lambda} \boldsymbol{W}_k(1)  
\\
&\equiv \boldsymbol{\Lambda} \big[ \boldsymbol{W}_k(r) - r   \boldsymbol{W}_k(1) \big].
\end{align*}
Consider the matrix moment defined by $\widehat{\boldsymbol{C}} = T^{-2} \sum_{t=1}^T \widehat{S}_t \widehat{S}_t$. Then an application of the CMT gives
\begin{align}
\widehat{\boldsymbol{C}} \Rightarrow \boldsymbol{\Lambda} \left\{ \int_0^1 \big[ \boldsymbol{W}_k(r) - r \boldsymbol{W}_k(1) \big]  \big[ \boldsymbol{W}_k(r) - r   \boldsymbol{W}_k(1) \big]^{\prime} dr \right\} \boldsymbol{\Lambda}^{\prime}
\end{align} 
Therefore, we denote with 
\begin{align}
\boldsymbol{P}_k = \int_0^1 \big[ \boldsymbol{W}_k(r) - r   \boldsymbol{W}_k(1) \big]  \big[ \boldsymbol{W}_k(r) - r   \boldsymbol{W}_k(1) \big]^{\prime} dr
\end{align}
which is the integral of the outer product of a $k-$dimensional multivariate Brownian bridge. In particular, in the univariate case $P_1$ is the limiting distribution of the Cramer-von Mises statistic and is related to the Anderson-Darling statistic. Furthermore, since $\boldsymbol{P}_k$ is positive-definite by construction, we can use a Cholesky decomposition to write $\boldsymbol{P}_k = \boldsymbol{Z}_k \boldsymbol{Z}_k$ or equivalently $\boldsymbol{P}_k^{-1} = \left( \boldsymbol{Z}_k^{\prime} \right)^{-1} \boldsymbol{Z}_k^{-1}$ where $\boldsymbol{Z}_k$ is lower triangular.

\bigskip

\begin{theorem}
If $y_t$ is generated by the location model and Assumptions hold, then
\begin{align}
\underset{ c \in \mathbb{R} }{ \mathsf{sup} } \left| \mathbb{P} \left( \frac{T \left( \widehat{\beta}_T - \beta \right)^2 }{2}  \leq c \right) - \mathbb{P} \left( \frac{ \mathcal{Z}^2 }{ \displaystyle \int_0^1 \kappa(r) B(r)^2 dr } \leq c \right) \right| = \mathcal{O} \left( T^{-1} \mathsf{log} T \right),
\end{align}
where $B(.)$ is a Brownian bridge and $\mathcal{Z} \sim \mathcal{N} (0,1)$ is independent of $B(.)$.
\end{theorem}

\end{wrap}

%%-------------------------------------------------------------------------%%
\newpage 

\bibliographystyle{apalike}

{\small 
\bibliography{myreferences1}}

\end{document}